\documentstyle[eqsecnum,aps,pra]{revtex}
\begin{document}
\twocolumn[\hsize\textwidth\columnwidth\hsize\csname
@twocolumnfalse\endcsname
\draft
\title{Non--Markovian Quantum Fluctuations and Superradiance Near a
Photonic Band Edge}
\author{Nipun Vats and Sajeev John}
\address{Department of Physics, University of Toronto, 60 St. George
Street, Toronto, Ontario, Canada M5S 1A7}
\date{June 9, 1998}
\maketitle

\widetext
\begin{abstract}
We discuss a point model for the collective emission of light from $N$
two-level atoms in a photonic bandgap material, each with an atomic resonant
frequency near the edge of the gap. In the limit of a low initial
occupation of the excited atomic state, our
system is shown to possess novel atomic spectra and population
statistics. For a high initial excited state population, mean
field theory suggests a
fractionalized inversion and a macroscopic 
polarization for the atoms in the steady state, both of which can be
controlled
by an external d.c.
field. This atomic steady state is accompanied by a non--zero
expectation value of the
electric field operators for field modes located in the vicinity of the
atoms. The nature of homogeneous broadening near the band edge is shown to
differ markedly from that in free space due to non--Markovian memory
effects in the
radiation dynamics.  Non--Markovian vacuum fluctuations are shown
to yield a partially coherent steady state polarization with a random phase. In
contrast with the steady state of a conventional laser, near a photonic band
edge this coherence occurs as a consequence of photon localization in
the absence of a
conventional cavity mode. We also introduce a classical stochastic function
with the same temporal correlations as the electromagnetic reservoir,
in order to stochastically simulate the effects of vacuum fluctuations
near a photonic band edge.       
\end{abstract}

\vfill
\narrowtext

\vskip2pc]

\section{Introduction}

In recent years, photonic bandgap (PBG) structures have been shown to lead
to the localization of light\cite{sjprl} through the carefully engineered
interplay between microscopic scattering resonances and the coherent
interference of light from many such scatterers\cite{erice}. Since the
initial proposal of photonic bandgaps\cite{{yablonovich},{sajeev}}, PBG
materials exhibiting photon localization have been fabricated at microwave
frequencies\cite{microwave} and more recently, large scale two--dimensional
PBG systems have been produced in the near-infrared\cite{ir}. The ultimate
goal for laser applications is a full three-dimensional PBG at optical
frequencies\cite{opbg1,opbg2,opbg3,opbg4}. A PBG comprises a range of
frequencies
over which linear photon propagation is prohibited. Therefore, atoms with
transition frequencies within the gap do not experience the usual
fluctuations in the electromagnetic vacuum that are responsible for
spontaneous decay. Instead, a photon--atom bound state is formed\cite
{johnwang}. Unlike the suppression of spontaneous emission from an atom in a
high--Q optical microcavity\cite{microcavity}, the bound photon may tunnel
many optical wavelengths away from the atom before being re--absorbed. Near
a photonic band edge, the photon density of states is
rapidly--varying, making
it dramatically different from the $\omega ^2$ dependence found in free
space. This implies that the nature of vacuum fluctuations and thus of
spontaneous emission near a band edge is radically different from that
of the exponential decay found in free space\cite{louisell}. More
fundamentally,
the correlation time of the electromagnetic vacuum fluctuations near a band
edge is not negligibly small on the time scale of the evolution of an atomic
system coupled to the electromagnetic field. In fact, the reservoir exhibits
long--range temporal correlations, making the temporal distinction between
atomic system and electromagnetic reservoir unclear. This renders the usual
Born--Markov approximation scheme invalid for band edge systems. Studies of
single atom spontaneous emission near a photonic band edge\cite
{{yeh},{sepaper}} have shown that this non--Markovian system--reservoir
interaction gives rise to novel phenomena, such as oscillatory behavior and
a fractional steady state population for a single excited atomic state, as
well as vacuum Rabi splitting and a sub--natural linewidth for atomic
emission.

We consider the Dicke model\cite{{dicke},{haroche}} for the collective
emission of light, or superradiance, from N identical two--level atoms
with a transition 
frequency near a photonic band edge. The study of superradiant emission is
of interest not only in its own right, but also because it provides a
valuable paradigm for understanding the self-organization and emission
properties of a band edge laser. Of late, there has been a resurgence of
interest in superradiance in the context of superradiant lasing action\cite
{haakelaser}, and due to the experimental realization of a true
Dicke superradiant system using laser--cooled atoms\cite{lasercool}. A low
threshold microlaser operating near a photonic band edge may exhibit unusual
dynamical, spectral and statistical properties. We will show that such
effects are already evident in band edge collective spontaneous emission. A
preliminary study of band edge superradiance for atoms
resonant with the band edge\cite{srpaper} has shown that
for an atomic system prepared initially with a small
collective atomic polarization, a 
fraction of the superradiant emission remains in the vicinity of the atoms,
and a macroscopic polarization emerges in the collective atomic steady
state. In addition to this form of spontaneous symmetry breaking, it has
been demonstrated that superradiant emission can proceed more quickly and
with greater intensity near a photonic band edge than in free
space. In the absence of an 
initial atomic polarization, the early stages of superradiance are governed
by fluctuations in the electromagnetic vacuum near the band edge. These
fluctuations affect the dynamics of collective decay and will determine the
quantum limit of the linewidth of a laser operating near a photonic band
edge.

The organization of the paper is as follows. In Section II, we present the
quantum Langevin equations for collective atomic dynamics in
band edge superradiance. In Section III, we calculate an approximate,
analytic solution
for the equations that describe the N--atom system with
low initial inversion of the atomic population. We show that the atoms can
exhibit novel
emission spectra and a suppression of population fluctuations near a band
edge.  Sections IV and V treat the
case of high initial inversion.  In Section IV, the mean field results
of Ref. \cite{srpaper} are extended to the case of
atoms with resonant
frequencies
displaced from the band edge. It is shown that the phase and amplitude of
the collective atomic polarization can be controlled by an external field
that Stark shifts the atomic transition relative to the band edge. The
dissipative effect of dipole dephasing is also included in the framework of our
non-Markovian system. 
Section V describes superradiant emission under the
influence of vacuum fluctuations by exploiting the temporal division of
superradiance into quantum and semi-classical regimes. We find that the
system exhibits a macroscopic steady state polarization amplitude with a
phase precession triggered by band edge quantum fluctuations. In Section
VI, we describe a method for generating a classical stochastic
function that simulates the effect of band edge vacuum fluctuations.
We show that, for a sufficiently large number of atoms, this classical
noise ansatz agrees well with the more exact simulations of Section V,
and may thus be useful in the analysis of band edge atom--field dynamics.
In Appendix A, We give the details of the calculation of the electromagnetic reservoir's
temporal autocorrelation function for different models of the photonic
band edge. This
correlation function is central to determining the nature of atomic decay.

\section{Equations of Motion}

We consider a model consisting of $N$ two--level atoms with a transition
frequency near the band edge coupled to the multi--mode radiation field in a
PBG material. For simplicity, we assume a point interaction, that is, the
spatial extent of the active region of the PBG material is less than the
wavelength of the emitted radiation. This is often referred to as the small
sample limit of superradiance\cite{haroche}. We neglect the
spatially random resonance dipole--dipole interaction (RDDI) near the band
edge, which may have a more important impact on atomic 
dynamics when the atomic transition lies deep within
the PBG\cite{{srpaper},{boseglass}}. Nevertheless, our simplified model
should provide a good qualitative picture of band edge collective emission.
For an excited atomic state $\left| 2\right\rangle $ and ground state $%
\left| 1\right\rangle $, the interaction Hamiltonian for our system can be
written as 
\begin{equation}
{\cal H}=\sum_\lambda \hbar \Delta _\lambda a_\lambda ^{\dag }a_\lambda
+i\hbar \sum_\lambda g_\lambda (a_\lambda ^{\dag }J_{12}-J_{21}a_\lambda ),
\label{ham}
\end{equation}
where $a_\lambda $ and $a_\lambda ^{\dag }$ are the radiation field
annihilation and creation operators respectively; $\Delta _\lambda =\omega
_\lambda -\omega _{21}$ is the detuning of the radiation mode frequency $%
\omega _\lambda $ from the atomic transition frequency $\omega
_{21}$.  
$g_\lambda =(\omega
_{21}d_{21}/\hbar )(\hbar /2\epsilon _0\omega _\lambda V)^{1/2}{\bf e}%
_\lambda \cdot {\bf u}_d$ is the atom-field coupling constant, where $d_{21}{\bf u}_d$ is the atomic dipole
moment vector, $V$ is the sample volume, and ${\bf e}_\lambda =e_{{\bf k}%
,\sigma },\;\sigma =1,2$ are the two transverse polarization vectors. The $%
J_{ij}$ are collective atomic operators, defined by the relation $%
J_{ij}\equiv \sum_{k=1}^N\left| i\right\rangle _{kk}\left\langle j\right|
;\;i,j=1,2$, where $\left| i\right\rangle _{k}$ denotes the $i$th
level of the $k$th atom. Using the Hamiltonian (\ref{ham}), we may write the Heisenberg
equations of motion for the operators of the field modes, $a_\lambda (t)$,
the atomic inversion, $J_3(t)\equiv J_{22}(t)-J_{11}(t)$, and the atomic
system's collective polarization, $J_{12}(t)$:

\begin{equation}
\frac d{dt}a_\lambda (t)=-i\Delta _\lambda a_\lambda (t)+g_\lambda J_{12}(t)
\label{la}
\end{equation}
\begin{equation}
\frac d{dt}J_3(t)=-2\sum_\lambda g_\lambda J_{21}(t)a_\lambda (t)+adj.
\label{lj3}
\end{equation}
\begin{equation}
\frac d{dt}J_{12}(t)=\sum_\lambda g_\lambda J_3(t)a_\lambda (t).
\label{lj12}
\end{equation}

We may adiabatically eliminate the field operators by formally integrating
equation (\ref{la}) and substituting the result into equations (\ref{lj3})
and (\ref{lj12}). The equations of motion for the collective atomic
operators are then 
\begin{eqnarray}
\frac d{dt}J_3(t)=&-&2\int_0^tJ_{21}(t)J_{12}(t^{^{\prime }})G(t-t^{^{\prime
}})dt^{^{\prime }} \nonumber \\
&-& 2J_{21}(t)\eta (t)+adj.  \label{j3eq}
\end{eqnarray}

\begin{equation}
\frac d{dt}J_{12}(t)=\int_0^tJ_3(t)J_{12}(t^{^{\prime }})G(t-t^{^{\prime
}})dt^{^{\prime }}+J_3(t)\eta (t).  \label{j21eq}
\end{equation}
Here, $\eta (t)=\sum_\lambda g_\lambda a_\lambda (0)e^{-i\Delta _\lambda t}$ is a
quantum noise operator which contains the influence of vacuum fluctuations. $%
G(t-t^{^{\prime }})$ is the time delay Green function, or memory kernel,
describing the electromagnetic reservoir's average effect on the time evolution
of the system operators. The Green function is given by the temporal
autocorrelation of the reservoir noise operator, 
\begin{equation}
G(t-t^{^{\prime }}) \equiv \left\langle \eta (t)\eta ^{\dag }(t^{\prime
})\right\rangle = \sum_\lambda g_\lambda ^2e^{-i\Delta _\lambda
(t-t^{\prime })}.
\label{gsum}
\end{equation}
We have made use of the fact that $\left\langle a_\lambda ^{\dag
}(0)a_{\lambda } (0)\right\rangle \simeq 0$, as we are dealing
with atomic
transition frequencies in the optical domain\cite{louisell}. In essence, $%
G(t-t^{^{\prime }})$ is a measure of the reservoir's memory of its previous
state on the time scale for the evolution of the atomic system. In
free space, the
density of field modes as a function of frequency is broad and slowly
varying, resulting in a Green
function that exhibits Markovian behavior, $G(t-t^{^{\prime }})=\frac \gamma 
2\delta (t-t^{^{\prime }})$, where $\gamma $ is the usual decay rate for
spontaneous emission\cite{louisell}. Near a photonic band edge, the density
of electromagnetic modes varies rapidly with frequency in a manner
determined by the photon dispersion relation, $\omega _{{\bf k}}$. We show
that this results in long range temporal correlations in the reservoir
which affect the
nature of the atom--field interaction.

In order to evaluate $G(t-t^{^{\prime }})$ near a band edge, we first
make the continuum approximation for the field mode sum in equation
(\ref{gsum}): 
\begin{equation}
G(t-t^{^{\prime }})=\frac{\omega _{21}^2d_{21}^2}{2\hbar \epsilon _0(2\pi
)^3 }\int \frac{d^3 {\bf k}}{\omega _{{\bf k}}}e_{}^{-i(\omega _{%
{\bf k}}-\omega _{21})(t-t^{^{\prime }})}.  \label{gint}
\end{equation}
In this paper, we use an effective                        
mass approximation to the full dispersion relation for a
photonic crystal.  Within this approximation, we
consider two models for
the near--band edge dispersion. The details of the calculation of
$G(t-t^{^{\prime }})$ 
for each model and a discussion of its applicability is given in 
Appendix A.  In an anisotropic dispersion model, appropriate to
fabricated PBG materials, we associate the band edge with a specific
point in $k$--space, ${\bf k}={\bf k}_0$.  By preserving the vector
character of the dispersion expanded about ${\bf k}_0$, we account for
the fact that, as ${\bf k}$ moves away from ${\bf k}_0 $, both the
direction and magnitude of the band edge wavevector are modified.
This gives a dispersion relation of the form:  
\begin{equation}
\omega _{{\bf k}}=\omega _c\pm A({\bf k}-{\bf k}_0)^2.  \label{dispersion}
\end{equation}
Here, $A=2c^2/\omega _{gap}$,
where $\omega _{gap}$ is the frequency width of the gap. The positive
(negative) sign indicates that $\omega _{{\bf k}}$ is expanded about the
upper (lower) edge of the PBG, and $\omega _c$ is the frequency of the
corresponding band edge. This form of dispersion is valid for a gap
width $\omega _{gap}\gg c\left| {\bf k}-{\bf k}_0\right| $, meaning that the
effective mass relation is most directly applicable to large photonic gaps
and for wavevectors near the band edge. Furthermore, for a large gap
and a collection of atoms which are nearly resonant with the upper
band edge, it is a very good approximation to completely neglect the
effects of the lower photon bands.  The band edge density of states
corresponding to equation (\ref{dispersion}) takes the form $\rho
(\omega)\sim (\omega -\omega _c)^{1/2},\,\omega >\omega _c$,
characteristic of a three--dimensional phase space.  The  resulting
Green function for $\omega_c(t-t^{^{\prime }})\gg 1$ is
\begin{equation}
G_A(t-t^{^{\prime }})=\frac{\beta _3^{1/2}e^{i[\pi /4+\delta
_c(t-t^{^{\prime }})]}}{(t-t^{^{\prime }})^{3/2}},\quad \,t>t^{^{\prime
}}.  \label{greenfn2}
\end{equation}

In addition to the anisotropic photon dispersion model, it is
instructive to consider a simpler isotropic model.  In this
model, we extrapolate the
dispersion relation for a 
one--dimensional gap to all three spatial dimensions.  We thus
assume that the Bragg condition is satisfied for the same wavevector
magnitude for all directions in $k$--space.  This yields an
effective mass dispersion of the form $\omega _k=\omega
_c+A(\left| {\bf k}\right| -\left| {\bf k}_0\right| )^2$, which associates
the band edge wavevector with a sphere in $k$--space, $\left| {\bf k}\right|
=k_0$.  Strictly speaking, an isotropic PBG at finite wavevector
$\left| {\bf k}_0 \right| $ does not occur in artificially created,
face centred cubic photonic crystals.  However, a nearly isotropic
gap near $k_0 = 0 $ occurs in certain polar crystals with
polaritonic excitations \cite{rupasov}.  A simple example of such a
crystal is table salt (NaCl), which has a polariton gap in the
infrared frequency regime.  The band edge density of
states in the isotropic model has the form $\rho
(\omega )\sim (\omega -\omega _c)^{-1/2},\,\omega >\omega _c$, the square
root singularity being characteristic of a one--dimensional phase
space.  For the Green function we obtain (see Appendix A), 
\begin{equation}
G_I(t-t^{^{\prime }})=\frac{\beta _1^{3/2}e^{-i[\pi /4-\delta
_c(t-t^{^{\prime }})]}}{(t-t^{^{\prime }})^{1/2}},\quad \,t>t^{^{\prime
}}.  \label{greenfn1}
\end{equation}

In both (\ref{greenfn2}) and (\ref{greenfn1}), $\delta_c
=\omega_{21}-\omega _c$ is the detuning of the atomic resonance
frequency from the
band edge, and $\beta_{\alpha} $ is a constant that depends on the
dimension of the
band edge singularity. In particular, for the isotropic model, $%
\beta _1^{3/2}= \omega _{21}^{7/2}d_{21}^2/12\hbar \epsilon _0\pi
^{3/2}c^3$,
while in the anisotropic model, $\beta _3^{1/2}=\omega
_{21}^2d_{21}^2/8\hbar \epsilon _0\omega _c\left( A\pi \right) ^{3/2}$.

\section{Low atomic excitation: harmonic oscillator model}

In order to understand the effects of band edge vacuum fluctuations, we begin
by presenting a simplified model that permits an analytic solution, and
is applicable to a system in which only a small
fraction of the two-level atoms are initially in their excited state.
This discussion demonstrates how light emission near a
photonic band edge can give rise to novel atomic dynamics, emission
spectra, and photon number statistics.  We write the atomic operators in
the Schwinger boson representation\cite{schwinger}: 
\begin{eqnarray}
J_{12}(t) &\rightarrow &b_1^{\dag }(t)b_2(t) \\
J_3(t) &\rightarrow &b_2^{\dag }(t)b_2(t)-b_1^{\dag }(t)b_1(t),
\end{eqnarray}
subject to the constraint on the total number of atoms, $b_1^{\dag
}(t)b_1(t)+b_2^{\dag }(t)b_2(t)=N$. The operators $b_i^{\dag }(t)$ and
$b_i(t)$ %
then describe transitions of the system between the excited
state ($i=2$) and the ground state ($i=1$). In the limit of low atomic
excitation, the state $\left| 1\right\rangle $ has a large population at all
times, meaning that we can replace the inversion operator by the classical
value $J_3(t)\approx -N$, and that $b_1(t)$ can be approximated by $%
b_1(t)\approx \sqrt{N}$. In this case, the initially excited
two--level atoms behave like a simple harmonic oscillator coupled to the
non--Markovian electromagnetic reservoir. A form of non-Markovian
coupling similar to that of bosons to the electromagnetic field occurs
in the context of the output coupling of a cold atom Bose
condensate from a trapping potential to the propagating modes of an
atom laser\cite{Hope}.  This mathematical analogy may lead to deeper
insight into both the atom laser problem and photonic band edge dynamics.
In our model, the Heisenberg equations of motion
(\ref{j3eq}) and (\ref{j21eq}), reduce to 
\begin{equation}
\frac d{dt}b_2(t)=-N\int_0^tb_2(t^{^{\prime }})G(t-t^{^{\prime
}})dt^{^{\prime }}-\sqrt{N}\eta (t).  \label{shoeq}
\end{equation}
Using the method of Laplace transforms, we can solve for $b_2(t) $ and find
\begin{equation}
b_2(t)=B(t)b_2(0)-\sqrt{N}\sum_\lambda A_\lambda (t)a_\lambda (0),
\label{shosoln}
\end{equation}
where 
\begin{eqnarray}
B(t) &=&{\cal L}^{-1}\left\{ \tilde B(s)\right\}, \label{bt} \\
\tilde B(s) &=& \left[ s+\tilde G(s)\right]^{-1},  \label{b} 
\end{eqnarray}
and
\begin{equation}
A_\lambda (t) ={\cal L}^{-1}\left\{ \frac{g_\lambda }{s+i\Delta _\lambda }%
\tilde B(s)\right\} . \label{a}
\end{equation}
${\cal L}^{-1}$ denotes the inverse Laplace transformation, and
$\tilde G(s)$  is the Laplace transform of the general memory kernel,
$G(t-t^{\prime}) $.  In this section, we consider the case of an
isotropic band edge in the effective mass approximation (equation
(\ref{greenfn2})), for which $\tilde G(s) $ is written as:
\begin{equation}
\tilde G_I(s)=\frac{N\beta _1^{3/2}e^{-i\pi /4}}{\sqrt{s-i\delta _c}}.
\label{gs}
\end{equation}
For this isotropic Green function, we denote the inverse Laplace
transform of equation (\ref{bt}) by $B_I(t) $. $B_I(t) $ was
computed in Ref.\cite{sepaper} in the context of single atom
spontaneous emission, and a detailed mathematical derivation may be
found therein. Here, it describes the mean
or drift evolution of our Heisenberg operator $b_2\left( t\right) $. The
solution has the form 
\begin{eqnarray}
B_I(t)=&2&a_1x_1e^{\beta _1x_1^2t+i\delta _ct}+a_2(x_2+y_2)e^{\beta
_1x_2^2t+i\delta _ct} \nonumber \\
&-&\sum_{j=1}^3a_jy_j\left[ 1-\Phi \left( \sqrt{\beta
_1x_j^2t}\right) \right] e_{}^{\beta _1x_j^2t+i\delta _ct}, \label{bsoln}
\end{eqnarray}
where 
\begin{equation}
x_1=(A_{+}+A_{-})e_{}^{i\pi /4},
\end{equation}
\begin{equation}
x_2=(A_{+}e^{-i\pi /6}-A_{-}e^{i\pi /6})e_{}^{-i\pi /4},
\end{equation}
\begin{equation}
x_3=(A_{+}e^{i\pi /6}-A_{-}e^{-i\pi /6})e_{}^{i3\pi /4},
\end{equation}
\begin{equation}
A_{\pm }=\left\{ \frac 12\pm \frac 12\left[ 1+\frac 4{27}\frac{\delta _c^3}{%
\beta _1^3}\right] ^{1/2}\right\} ^{1/3},
\end{equation}
\begin{equation}
y_j=\sqrt{x_j^2},\quad j=1,2,3.
\end{equation}
$\Phi \left( x\right) $ is the error function, $\Phi \left( x\right) =%
\frac 2{\sqrt{\pi }}\int_0^xe^{-t^2}dt$. 

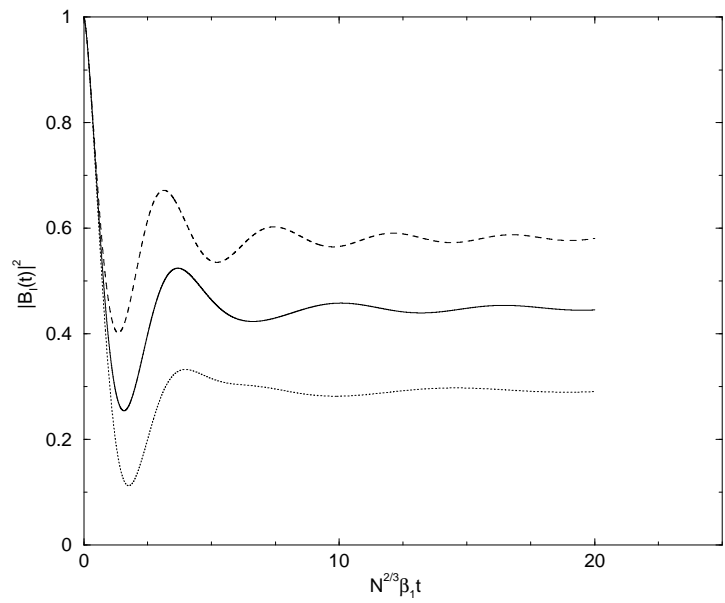
\begin{figure}[tbp]
\caption{Normalized population of the excited atomic state near an
isotropic photonic band edge for low initial
atomic excitation.  Various values of the detuning,
$\delta_c \equiv
\omega_{21} - \omega_c $, of the atomic resonant frequency
$\omega_{21} $ from a band edge at frequency
$\omega_c$ are shown. Dashed line,
$\delta_c = -.5$; solid line, $\delta_c = 0$;
dotted line, $\delta_c = .5$.  $\delta_c$ is measured
in units of $N^{2/3}\beta_1 $.}
\label{fig6}
\end{figure}

The probability of finding the atoms in the excited state is given
by $ \left\langle b_2^{\dag}(t) b_2(t) \right\rangle = 
\left| B_I(t)\right| ^2$, and is plotted in Fig. \ref{fig6}.  We find
that the excited state population
exhibits decay and
oscillatory behavior before reaching a non-zero steady state value due to
photon localization. These effects are due to the strong dressing of
the atoms by the radiation field near a photonic band edge, resulting
in dressed atomic states that straddle the band edge.  Light emission
from the dressed state outside
the gap results in highly non-Markovian decay of the atomic
population, while
the dressed state  shifted into the gap is
responsible for the fractional steady state population of the excited
state.  The consequences of this strong atom--field interaction are
discussed in detail for single atom spontaneous emission in
Ref. \cite{sepaper}, and for superradiant emission in Sections IV and V
of this paper.  We
note that the
degree of steady state localization is a sensitive
function of the detuning, $\delta_c$, of the atomic resonance from the
band edge.  The decay rate scales as $N^{2/3}\beta _1t$
for the isotropic
model.  However, there is no
evidence for the build-up of inter--atomic coherence, as very few of the atoms
are initially excited. 

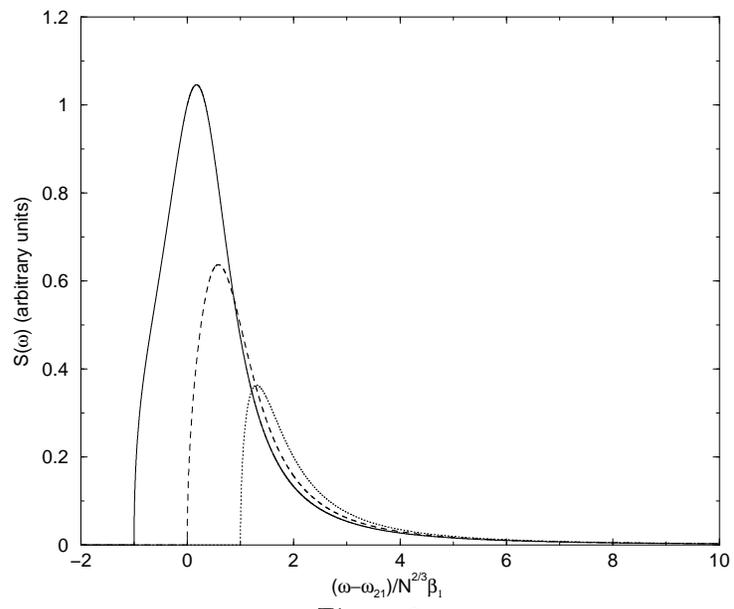
\begin{figure}[tbp]
\caption{Collective atomic emission spectrum ${\cal S}(\omega )$
(arbitrary units) near an
isotropic band
edge for low initial atomic excitation.  Various values of the
detuning, $\delta_c \equiv \omega_{21} - \omega_c $, of the atomic
resonant frequency $\omega_{21} $ from an isotropic photonic band edge
at frequency
$\omega_c$ are shown.  Dotted line,  $\delta_c = -1$; dashed line,  $\delta_c
5=0$; solid line,  $\delta_c = 1$. $\delta_c$ is measured in units of
$N^{2/3}\beta_1 $.}
\label{fig7}
\end{figure}

Equation (\ref{shoeq}) also allows us to calculate
the system's emission spectrum into the modes $\omega $ for an atom
with resonant frequency $\omega_{21} $ using the relation 
\begin{eqnarray}
{\cal S}\left( \omega \right) &=& \int_{0
}^\infty e^{-i\left( \omega - \omega_{21}\right) \tau }\left\langle
b_2^{\dag }\left( \tau \right)
b_2\left( 0\right) \right\rangle d\tau + c.c. \nonumber \\
&\sim& Re \left
\{ \tilde B^* \left[ i \left( \omega - \omega_{21}\right)\right] \right \},
\end{eqnarray}
where  $\tilde B(s) $ is defined in equation (\ref{b}).  The spectrum
for the isotropic model
is then
\begin{equation}
{\cal S}_I\left( \omega \right) \simeq \left\{ 
\begin{array}{c}
\qquad \qquad \qquad 0 \qquad \qquad \quad \quad ,\quad \omega \leq
\omega_c \\ 
N\beta _1^{3/2}\frac{\sqrt{\omega-\omega_c}}{N^2\beta_1 ^3+\left
( \omega-\omega_{21}\right)^2 \left( \omega-\omega_c \right)}\quad,\quad
\omega > \omega_c
\end{array}
\right.
\end{equation}
This spectrum is shown in Fig. \ref{fig7}, and differs significantly from
the Lorentzian spectrum for light emission in free
space. In fact, the emission spectrum is not centered about the atomic
resonant frequency, which is what one would expect for an atom
decaying to an unrestricted vacuum mode density.  We see that for an
arbitrary detuning, $ \delta_c
$, of $\omega_{21} $ from
the band edge, the emission spectrum vanishes for frequencies at the
band edge and within the gap, $ \omega \leq \omega_c $.  This is
consistent with the
localization of light near the atoms for electromagnetic modes within
the PBG. As $ \omega_{21} $ is detuned further into the gap, spectral
results confirm that a greater fraction of the light is localized in
the gap dressed state, as the total emission intensity out of the
decaying dressed state is reduced. Conversely, as $\omega_{21} $ is
moved out of the gap, the emission profile becomes closer to a
Lorentzian in form and the total emitted intensity increases.  The spectral
linewidth ratio between the isotropic band edge and free space is of the
order of $\beta _1/(\gamma N^{1/3})$, while for an anisotropic band edge it
is $\sim N\beta _3/\gamma $. This corresponds to the fact that
collective emission is much more
rapid near an anisotropic band edge than in free space, whereas it is
slower than in free space for the isotropic model.

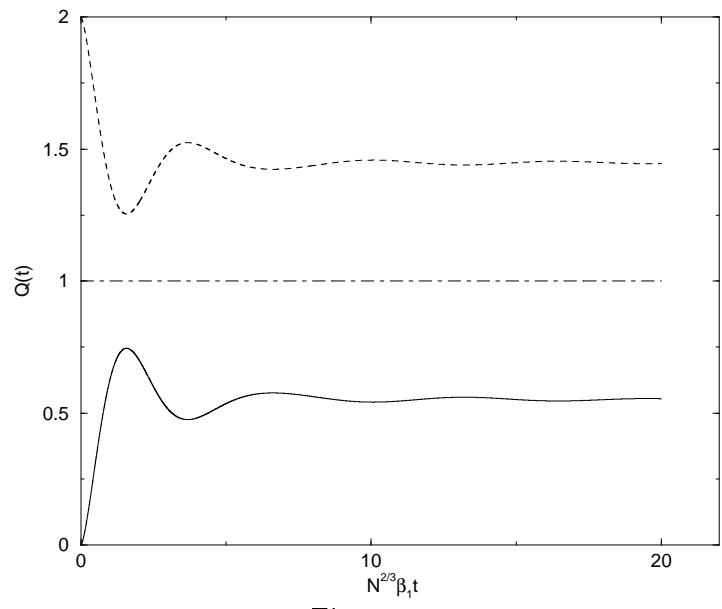
\begin{figure}[tbp]
\caption{Fluctuations in the excited state atomic population as
measured by the Mandel parameter, $Q(t) =
(\left\langle n^2(t)\right\rangle -\left\langle n(t)\right\rangle ^2) /
\left\langle n(t)\right\rangle$, for low initial excitation for an
atomic resonant frequency tuned to an isotropic photonic band edge,
$\delta_c = 0$.
Dashed line, $Q(0)=2$; solid line, $Q(0)=0$. Long-short dashed line denotes
fluctuations for Poissonian population variance, $Q(0)=1$.}
\label{fig8}
\end{figure}

It is also instructive to evaluate the quantum fluctuations in the atomic
inversion in the context of the harmonic oscillator model.  Variances
in the atomic population can be written in terms of the Mandel
Q-parameter\cite
{mandel}, 
\begin{equation}
Q(t)=\frac{\left\langle n^2(t)\right\rangle -\left\langle n(t)\right\rangle
^2}{\left\langle n(t)\right\rangle },  \label{mandel}
\end{equation}
where $n(t)\equiv b_2^{\dag }(t)b_2(t)$ is the number operator for the
occupation of the excited state. Since both the free space and PBG solutions
in our model can be written in the form of equation (\ref{shosoln}), we can
write the Q-parameter in the general form
\begin{equation}
Q(t)=\left| B(t) \right|^2 Q(0)+ N \sum_{\lambda}\left|A_{\lambda}
(t)\right|^2.   \label{sj1}
\end{equation}
Again, $\left| B(t)\right| ^2$ is the normalized probability of finding the
initially excited fraction of the atoms still in the excited state at
time $t$.  For an isotropic band edge, $B(t)=B_I(t)$ (equation
(\ref{bsoln})), whereas in free space, $%
B(t)\sim e^{-N\gamma t/2}$, representing the exponential decay of the
excited state population.  Using  the identity $N\sum_\lambda \left| A_\lambda
(t)\right| ^2= 1-\left| B\left( t\right)
\right| ^2$, as derived in Appendix B, we can write the population fluctuations
as  
\begin{equation}
Q(t)=\left| B(t)\right| ^2\left[ Q(0)-1\right] +1.  \label{q}
\end{equation}
For arbitrary initial statistics, atoms in free
space decay to the vacuum state with $Q\left( t\right) =1$; since the atoms
decay fully, there are no meaningful atomic statistics in the long time
limit. $Q\left( t\right) $ is plotted in Fig. \ref{fig8} for the isotropic
band edge ($\delta _c=0$) for the cases $Q(0)=0$, $1$, and $2$. Near the
band edge, photon localization prevents the atomic system from decaying to
the ground state. We find instead that the steady state statistics
are sensitive to the statistics of the initial state and to the value of $%
\delta _c$. A system initially prepared with super--Poissonian statistics ($%
Q(0)>1$) experiences a {\em suppression} of population fluctuations in the
steady state. In a system that is initially sub--Poissonian ($Q(0)<1$), the
fluctuations increase, but are held below the Poissonian level by photon
localization. In both cases, the steady state value of the atomic population
fluctuations is controlled by $ \delta_c $. Our harmonic oscillator
model thus suggests that
a PBG system may exhibit novel quantum statistics in the absence of a cavity
or external fields.
It is important to extend the analysis of
collective emission under the influence of vacuum fluctuations to the
high excitation (superradiant) regime.  In this case, the two--level
nature of the atomic operators will become important and will modify the
quantum statistics from that of the harmonic oscillator picture.  This
generalization is considered in the next two sections.

\section{High atomic excitation: Mean Field Solution}

When the atomic system is initially fully or nearly fully inverted, we
expect inter--atomic coherences, transmitted via the atomic
polarizations, to have a
strong influence on emission dynamics.  For such high initial atomic
excitation, the quantum Langevin equations (\ref{j3eq}) and
(\ref{j21eq}), paired with 
the non--Markovian memory kernels (\ref{greenfn2}) or (\ref{greenfn1}),
 do not possess
an obvious analytic solution. Moreover, conventional perturbation theory
applied to these equations fails to recapture the influence of the
photon--atom bound state\cite{johnwang}, which plays a crucial role in band
edge radiation dynamics. However, when the superradiant system is
prepared with a
non-zero initial polarization ($J_{12}(0)\neq 0$), the
average dipole moment dominates the incoherent effect of the vacuum
fluctuations and the subsequent evolution is well--described by a
semi--classical approximation\cite{haroche}. In this case, it is possible to
factorize the atomic operator equations: 
\begin{equation}
\frac d{dt}\left\langle J_3(t)\right\rangle =-4Re\left\{ \left\langle
J_{21}(t)\right\rangle \int_0^t\left\langle J_{12}(t^{^{\prime
}})\right\rangle G(t-t^{^{\prime }})dt^{^{\prime }}\right\}   \label{semij3}
\end{equation}
\begin{equation}
\frac d{dt}\left\langle J_{12}(t)\right\rangle =\left\langle
J_3(t)\right\rangle \int_0^t\left\langle J_{12}(t^{^{\prime }})\right\rangle
G(t-t^{^{\prime }})dt^{^{\prime }}.  \label{semij12}
\end{equation}
The brackets $\left\langle {\cal O}\right\rangle $ denote the quantum
mechanical average of the Heisenberg operator ${\cal O}$ over the Heisenberg
picture atom-field state vector, $\left| \Psi \right\rangle =\left|
vac\right\rangle \otimes \left| \psi \right\rangle $, where $\left|
vac\right\rangle $ represents the electromagnetic vacuum state, and $\left|
\psi \right\rangle $ represents the initial state of the atomic system.
Clearly in this mean field approach, the quantum noise contribution is
neglected, as $%
\left\langle \eta (t)\right\rangle =0$. Recently, Bay, Lambropoulos and
M\o lmer\cite{bay} found that, for
a simpler Fano profile gap model, the dynamics of superradiant emission
are affected by the choice of factorization
applied to the full quantum equations. However, the complete factorization
used here retains the qualitative features and evolution time scales of more
elaborate factorization schemes. Equations (\ref{semij3}) and (\ref{semij12}%
) were solved numerically in reference \cite{srpaper} for an atomic
resonance frequency coincident with the band edge ($\delta _c=0$) and a
small initial collective polarization. The initial collective state
was assumed to be of the form 
\begin{equation}
\left| \psi \right\rangle =\prod_{k=1}^N\left( \sqrt{r}\left|
1\right\rangle +\sqrt{1-r}\left| 2\right\rangle \right) _k
\label{istate}
\end{equation}
with $r\ll 1$, so that initially the atoms are almost fully inverted. In
this paper, we extend the previous analysis to atomic frequencies detuned
from the band edge. Despite its neglect of vacuum fluctuations, mean
field theory
illuminates many of the interesting
features of the system.  The relationship between mean field theory and a more
complete description including quantum fluctuations is discussed in Section
V.  

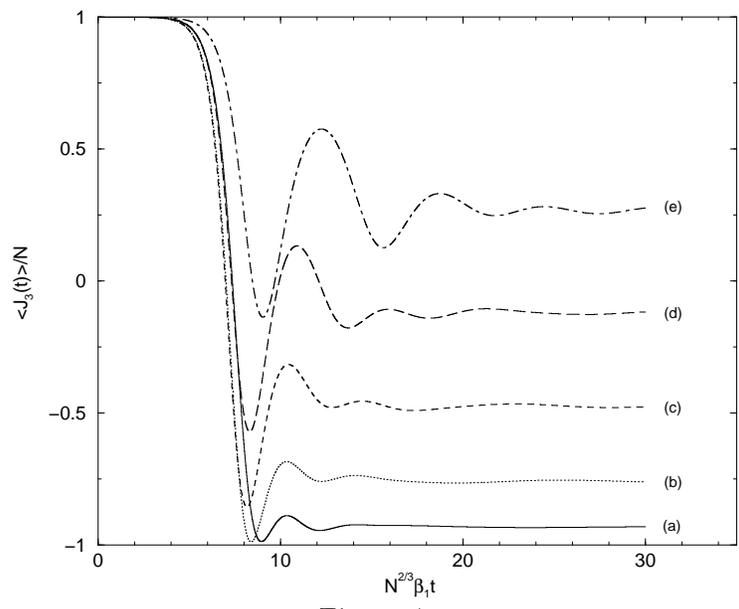
\begin{figure}[tbp]
\caption{Mean field solution for the atomic inversion, $\left\langle
J_3(t)\right\rangle /N$, near an
isotropic photonic band edge, starting with an infinitesimal initial
polarization, $r=10^{-5} $.  Various values of the
detuning, $\delta_c \equiv \omega_{21} - \omega_c $, of the atomic
resonant frequency $\omega_{21} $ from a band edge
at frequency
$\omega_c$ are shown. (a) $\delta
_c= 1$; (b) $\delta _c=.5 $; (c) $\delta _c=0$; (d) $\delta _c=-.5 $;
(e) $\delta _c=-1$. $\delta_c$ is measured in units of $N^{2/3}\beta_1 $.}
\label{fig1}
\end{figure}

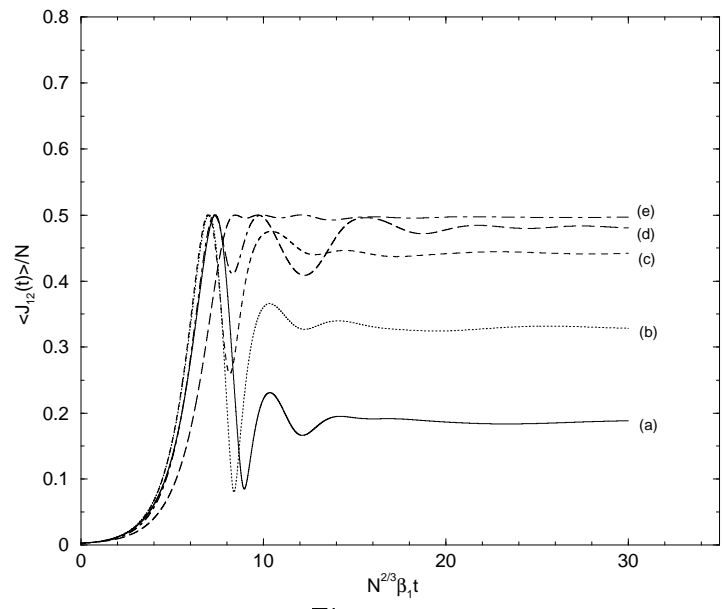
\begin{figure}[tbp]
\caption{Mean field solution for the atomic polarization amplitude,
$\left| \left\langle J_{12}(t)\right\rangle \right| /N $, near an
isotropic photonic band edge, starting with
an infinitesimal initial
polarization, $r=10^{-5} $.  Various values of the
detuning, $\delta_c \equiv \omega_{21} - \omega_c $, of the atomic
resonant frequency $\omega_{21} $ from a band edge
at frequency
$\omega_c$ are shown. (a) $ \delta _c=1 $; (b) $\delta _c=.5 $; (c) $\delta
_c=0$; (d) $\delta_c=-.5$; (e) $\delta _c=-1 $.  $\delta_c$
is measured in units of $N^{2/3}\beta_1 $.}
\label{fig2}
\end{figure}

For clarity, we discuss separately
the atomic dynamics in our isotropic and anisotropic dispersion
models.  Figures \ref{fig1} and \ref{fig2} show the
inversion per atom
and the average polarization amplitude per atom respectively for various
values of $\delta _c$ near an
isotropic band edge . We see from
Fig. \ref{fig1} that a fraction of the superradiant emission remains
localized in the
vicinity of the atoms in the steady state, due to the Bragg reflection of
collective radiative emission back to the atoms. This
localized light exhibits a non--zero expectation value for the field
operator, which in turn leads to the emergence of a macroscopic polarization
amplitude in the steady state.  We further note that the decay rate
for the upper atomic state  is proportional to $N^{2/3} $.
Accordingly, the peak 
radiation intensity is proportional to $N^{5/3} $.  This is to
be compared with the values $N $ and $N^2 $ for the free space decay
rate and peak radiation intensity respectively.  

As in single
atom spontaneous emission
near an isotropic band edge \cite
{sepaper}, the dressing of the atoms by their own radiation field causes a
splitting of the band of collective atomic states such that the
collective spectral
density vanishes at the band edge frequency. The strongly-dressed atomic
states are repelled from the band edge, with some levels being pulled
into the gap and the remaining levels being pushed
into the electromagnetic continuum outside the PBG. In the long time (steady
state) limit, the energy contained in the dressed states outside the
bandgap decays whereas the energy in the states
inside the gap remains in the vicinity of the emitting atoms. It is
the localized light associated with the gap
dressed states which sustains the fractionalized steady state inversion and
non-zero atomic polarization. For the isotropic model, this splitting and
fractional localization persist even when $\omega_{21} $ lies
just outside the gap ($\delta _c>0$), and the fraction of localized
light in the
steady state increases as $\omega _{21}$ moves towards and enters the gap. In
the dressed state picture, the self--induced oscillations in both the
inversion and the polarization which occur during radiative emission can be
interpreted as being due to interference between the dressed states. The
oscillation frequency is proportional to the frequency splitting between the
upper and lower collective dressed states. This is the analogue of the
collective Rabi oscillations of N Rydberg atoms in a resonant high-Q cavity 
\cite{rabipaper}. From Fig. \ref{fig1}, we see that a dressed state outside the
band gap decays more slowly for atomic resonant frequencies deeper inside
the gap, causing the collective oscillations to persist over longer periods
of time. Clearly, this decay is non-exponential and highly
non--Markovian in nature. Fig. \ref{fig2} confirms that, as required, the
polarization amplitude for large
negative values of $\delta _c$ is constrained by the condition, $%
\left\langle J_{12}(t)\right\rangle /N\leq 1/2$.

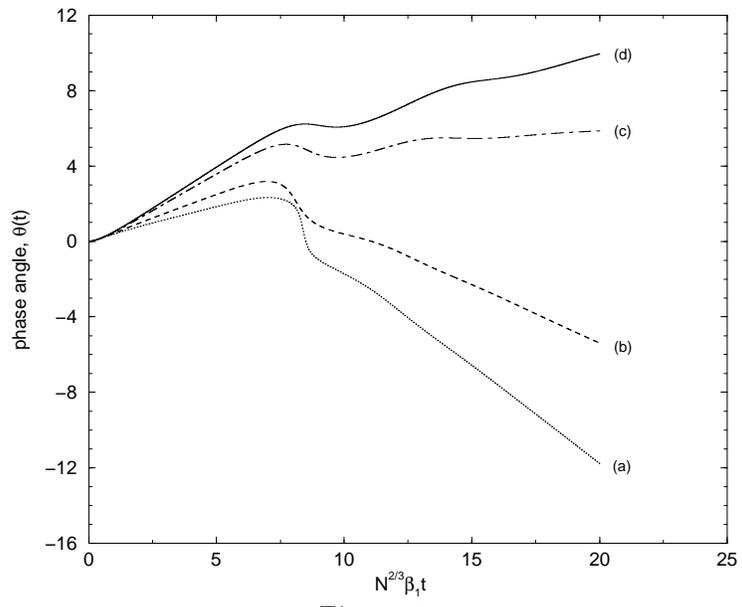
\begin{figure}[tbp]
\caption{Mean field solution for the phase angle of the atomic
polarization, $\theta (t)$, near an
isotropic photonic band edge, starting with
an infinitesimal initial
polarization, $r=10^{-5} $.  Various values of the
detuning, $\delta_c \equiv \omega_{21} - \omega_c $, of the atomic
resonant frequency $\omega_{21} $ from a band edge
at frequency
$\omega_c$ are shown. (a) $\delta _c=.5$; (b) $%
\delta _c=0$; (c) $\delta _c=-.75$; (d) $\delta _c=-1$. $\delta_c$
is measured in units of $N^{2/3}\beta_1 $.}
\label{fig3}
\end{figure}

In Fig. \ref{fig3}, we plot the phase angle of the collective
atomic polarization in the isotropic model,
$\theta(t)= tan^{-1}\left\{ Im \left\langle J_{12}(t)\right\rangle /
Re\left\langle J_{12}(t)\right\rangle\right\} $. Prior to atomic
emission, this
phase angle rotates at a constant rate, and in the vicinity of the decay
process $\theta (t)$ exhibits the effects of collective Rabi oscillations.
When the emission is complete, the rate of change of phase angle, $\dot 
\theta (t)$, attains a new steady state value, $\dot \theta (t_s)$, that
depends sensitively on the detuning frequency $\delta _c$. $\dot \theta (t_s)
$ is a measure of the energy difference between the bare atomic state and
the localized dressed state, $\hbar\left(\omega _{21}-\omega _{loc}\right)$.  Such a
polarization phase rotation
implies that the collective
atomic Bloch vector of the system exhibits precessional dynamics in the
steady state. Unlike the conventional precession\cite{eberlybk} of atomic
dipoles in an ordinary vacuum driven by an external laser field, Bloch
vector precession in a PBG occurs in the absence of an external driving
field.  Instead, the precession is driven by the self--organized
state of light generated by superradiance, which remains localized near the
emitting atoms. We see in Fig. \ref{fig3} that for values of $\delta _c$ such
that $\omega _{21}-\omega _{loc}<0$, $\dot \theta (t_s)$ is negative, while
for $\omega _{21}-\omega _{loc}>0$, $\dot \theta (t_s)$ is positive, i.e.
the phase is rotating in the opposite direction. At a detuning corresponding
to a constant phase in the steady state ($\dot \theta (t_s)=0$), the dressed
and bare states are of the same energy; this occurs for a detuning value of $%
\delta _c=-0.644 N^{2/3}\beta _1$. At this value of $\delta _c$, we also find that $%
\left\langle J_3(t_s)\right\rangle =0$, implying that there is no net
absorption of light by the atomic system. This is, in essence, a collective
transparent state\cite{eberlybk}.

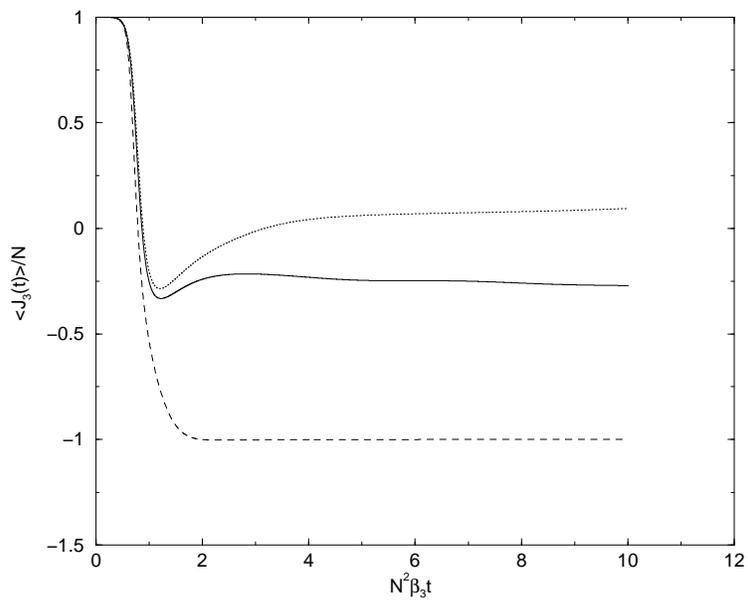
\begin{figure}[tbp]
\caption{Mean field solution for the atomic inversion, $\left\langle
J_3(t) \right\rangle /N$, near an
anisotropic photonic band edge, starting with
an infinitesimal initial
polarization, $r=10^{-6} $.  Various values of the
detuning, $\delta_c \equiv \omega_{21} - \omega_c $, of the atomic
resonant frequency $\omega_{21} $ from a band
edge at frequency
$\omega_c$ are shown.  Dashed
line, $\delta_c = .1 $; solid line, $\delta_c = 0 $; dotted line,
$\delta_c = -.3 $.  $\delta_c $ is measured in units of $N^2\beta_3 $.}
\label{fig4}
\end{figure}

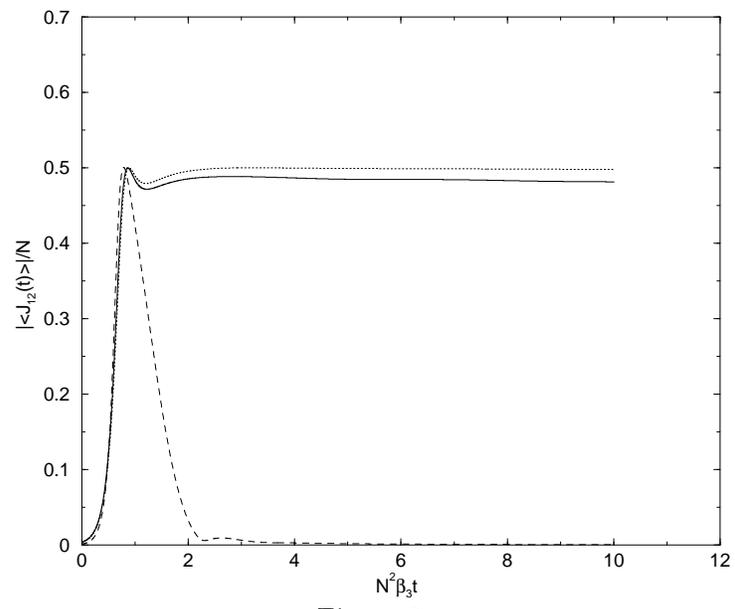
\begin{figure}[tbp]
\caption{Mean field solution for the atomic polarization amplitude, 
$\left| \left\langle J_{12}(t)
\right\rangle \right| /N $, near an
anisotropic photonic band edge, starting with
an infinitesimal initial
polarization, $r=10^{-6} $.  Various values of the
detuning, $\delta_c \equiv \omega_{21} - \omega_c $, of the atomic
resonant frequency $\omega_{21} $ from a band
edge at frequency 
$\omega_c$ are shown.  Dashed
line, $\delta_c = .1 $; solid line, $\delta_c = 0 $; dotted line,
$\delta_c = -.3 $.  $\delta_c $ is measured in units of $N^2\beta_3 $.}
\label{fig4b}
\end{figure}

Collective emission dynamics near an anisotropic band edge are
pictured in Figs. \ref{fig4} and \ref{fig4b}.  For $\omega_{21} $
coincident with the band edge or slightly within the gap ($\delta_c
\leq 0 $), we again find a fractional atomic inversion in the steady
state (Fig. \ref{fig4}).  Rabi oscillations in the atomic population
are much less pronounced than in the isotropic model, even for
$\omega_{21} $ detuned into the gap.  This demonstrates that the
dressed atomic states outside a physical photonic band edge decay much
more rapidly than the isotropic model would suggest.
Furthermore, in contrast with the isotropic model, we see that photon
localization is lost for even a small detuning of $\omega_{21} $ into
the continuum of field modes outside the band edge.  Therefore,
while we find a macroscopic steady state polarization and precessional
dynamics of the Bloch vector for $\delta_c
\le 0 $ (Fig. \ref{fig4b}), for $\delta_c > 0 $ the polarization dies
away after collective emission has taken place.  Photon localization
from an atomic level lying just outside the gap in a three dimensional
PBG material may, however, be realized through quantum interference
effects if there is a third atomic level lying slightly inside the
gap\cite{mesfin2}.  These results
point to the greater sensitivity of the atomic dynamics to the more realistic
anisotropic band edge.  Because
the isotropic model overestimates the momentum space for photons
satisfying the Bragg condition, photon localization effects and
vacuum Rabi splitting are exaggerated in the isotropic model relative
to an artificial photonic crystal.  In the anisotropic model, the
phase space available for propagation vanishes as the optical
frequency approaches the band edge.  As a result, vacuum Rabi
splitting pushes the collective atomic dressed state into a region
with a larger density of electromagnetic modes.  Consequently, the decay
rate of the atomic inversion is proportional to $N^2$ near the anisotropic
band edge, and the corresponding peak radiation intensity is
proportional to $N^3$. Clearly, superradiance
near an anisotropic PBG can proceed more quickly and can be more intense
than in free space. As a result, PBG superradiance may enable
the design of mirrorless, low--threshold microlasers exhibiting ultrafast
modulation speeds. A more complete analysis of lasing near a photonic band
edge including pumping and dissipative effects will be presented elsewhere.

From polarization phase and amplitude results, we conclude that: (i)
Unlike in free space, the atoms
near a photonic band edge attain a fractionally inverted state with constant
polarization amplitude and rate of change of phase angle. This corresponds
to a macroscopic atomic coherence in the steady state analogous to that
experienced in a laser. In our case however, ''lasing'' occurs in the
band edge continuum rather than into a conventional cavity mode. (ii)
By varying the value of $\delta
_c$, one may control the direction and rate of change of the steady state
polarization phase angle. This may be realized by applying a small external
d.c. field to the sample which Stark shifts the atomic transition frequency
of the atoms. This type of control over the collective atomic Bloch vector
may be of importance in the area of information storage and optical memory
devices\cite{memory,mesfin}.

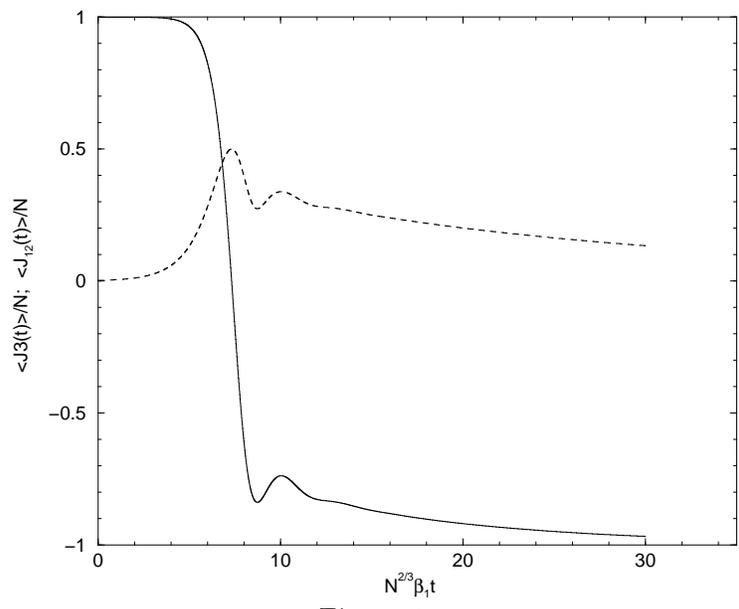
\begin{figure}[tbp]
\caption{Mean field solution for the atomic inversion (solid line) and
polarization amplitude (dashed line) under the influence of collision
broadening for an atomic resonant frequency at an isotropic photonic
band edge, $\delta_c = 0 $.  The system is given an infinitesimal
initial polarization, $r=10^{-5} $.  The simulated stark shift is a
Gaussian random distribution with zero mean and standard deviation $.5
N^{2/3}\beta_1 $.}
\label{fig5}
\end{figure}

The above analysis makes it clear that collective spontaneous
emission dynamics in a PBG are significantly different from those in free
space. In a real PBG material, the dephasing of atomic dipoles due to
interatomic collisions or phonon--atom interactions may also have a
significant effect on the evolution of our system over a large range of
temperatures. In the free space Markov approach, dipole dephasing is
described by a phenomenological polarization decay constant\cite{mands}.
Since the Markov approximation does not apply near a band edge, one cannot
account for dephasing by simply adding a phenomenological decay term to
equation (\ref{semij12}). However, we expect that the atomic resonant
frequency will experience random Stark shifts due to atom--atom or
atom--phonon interactions. This effect can be included in the description of
our system by adding a variation $\Delta $ to the detuning frequency $\delta
_c$ at each time step in a computational simulation of equations (\ref
{semij3}) and (\ref{semij12}). $\Delta $ is chosen to be a Gaussian random
number with zero mean. The width of the Gaussian distribution is determined
by the magnitude of the random Stark effect. Such a simulation in free space would
include a random $\Delta $ only in the equation for the atomic polarization.
This is because the slowly varying photon density of states seen by
the atoms at the frequency $%
\omega _{21}+\Delta $ does not change significantly with typical
homogeneous line
broadening effects. In contrast, we have seen that near a photonic band
edge, slight variations in $\delta _c$ may drastically change the atomic
inversion. Therefore we include $\Delta $ in both system equations. In Fig.
\ref{fig5}, we plot the evolution of the collective inversion and polarization under
the simulated collision broadening described above. The random Stark shifts
lead to the loss of macroscopic polarization and the loss of atomic
inversion in the long time limit. The latter effect can be understood by
noting that the random frequency shifts are symmetrically distributed about
the mean resonant frequency. Frequency shifts into the gap promote photon
localization, while those away from the gap cause further decay of the
atomic inversion. Over time, the net result is that the frequency shifts
away from the gap encourage the decay of the atomic population.  This
is true even in
atomic systems for which the mean resonant frequency lies within the gap.
From the above considerations, it is clear that dephasing is a significant
perturbation on photon localization near a photonic band edge. In
particular, dephasing determines the threshold external pumping
required to achieve atomic inversion in a band edge laser.  It also
facilitates the emission of laser light from the photonic crystal.

Although a superradiant system can be prepared in a coherent initial state
of the type described by equation (\ref{istate})\cite{eberlybk},
collective emission is typically
initiated by spontaneous emission, a random, incoherent process. Over time,
spontaneous emission leads to the build-up of macroscopic coherence in the
sample. The effect of vacuum fluctuations is then of considerable importance
in the full description of superradiance, both from a fundamental point of
view, and for potential device applications, such as the recently proposed
superradiant laser\cite{haakelaser}.  In the next section, we present
a more detailed description of PBG superradiance that takes into
account the role of quantum fluctuations.

\section{Band Edge Superradiance and Quantum Fluctuations}

In order to describe the evolution of the superradiant system's collective
Bloch vector under the influence of quantum fluctuations, we consider atomic
operator correlation functions of the form\cite{polder} 
\begin{equation}
g^{pq}=\left\langle (J_{12})^p(J_{21})^q\right\rangle .
\end{equation}
Here the operators are evaluated at equal times. As in free space, we expect
vacuum fluctuations to drive the system from its unstable initial state with
all atoms inverted to a new stable equilibrium state. Such
fluctuations are particularly relevant prior to the
build-up of macroscopic atomic polarization.  Indeed, they provide the
trigger for
superradiant emission. In the early--time, inverted regime, we may set
$J_3(t)=J_3(0)$ in
equations (\ref{j3eq}) and (\ref{j21eq}), giving 
\begin{equation}
\frac d{dt}J_{12}(t)=\int_0^t dt^{\prime}J_3(0)J_{12}(t^{^{\prime }})G(t-t^{^{\prime
}})+J_3(0)\eta (t).  \label{earlyeq}
\end{equation}
The resulting equation remains non-linear, and involves products of atomic
and reservoir operators. We may simplify expressions containing operators in
this inverted regime by considering operator averages over only the atomic
Hilbert space. For an arbitrary Heisenberg operator ${\cal O}(t)$, we denote
the atomic expectation value for an initial fully inverted state
$\left|I\right\rangle$ by $\left\langle {\cal O}\right\rangle _A\equiv
\left\langle I\right| {\cal O}\left| I\right\rangle $. We denote by
the set $\left\{
\left| \,\lambda \right\rangle \right\}$ a complete set of $2^N$
normalized basis vectors for the atomic Hilbert space including
$\left| I \right\rangle $, such that $\left\langle \lambda | I
\right\rangle = \delta_{\lambda , I} $, where $\delta _{\alpha
,\beta }$ is the Kronecker delta function.  Clearly,
$\,\left\langle \,I\,\right| \,J_3(0)\left|
\,\lambda \right\rangle =N\delta _{I,\lambda }$. Since $J_3(0)$ acts as a
source term for $J_{12}(t)$ in  equation (\ref{earlyeq}), we also have
the property $\left\langle I\,\right| \,J_{12}(t)\left|
\,\lambda \right\rangle =0$ for $\lambda \neq I $ in the inverted
regime.  This can be shown by considering the equation of motion for
$\left\langle I \right| J_{12}(t) \left| \lambda \right\rangle $:
\begin{eqnarray}
& \frac{d}{dt} & \left\langle I \right| J_{12}(t) \left| \lambda
\right\rangle \nonumber \\
&=& \int_0^t dt^{\prime} \sum_{\mu} \left\langle I
\right| J_3(0) \left| \mu \right\rangle \left\langle \mu \right|
J_{12}(t^{\prime}) \left| \lambda \right\rangle G(t-t^{\prime})
\nonumber \\
&\qquad& + 
\left\langle I \right| J_3(0) \left| \lambda \right\rangle \eta (t)
\nonumber \\
&=& N\int_0^t dt^{\prime} \left\langle I \right| J_{12}(t^{\prime})
\left| \lambda \right\rangle G(t-t^{\prime}) + N\delta_{I,\lambda
}\eta(t), \label{proof}
\end{eqnarray}
where $\mu $ labels a complete set of atomic states.  This
integro--differential equation satisfies the initial condition
$\left\langle I \right| J_{12}(0) \left| \lambda \right\rangle = 0 $,
since $ J_{12}(0) $ acts as a raising operator on the fully inverted
bra vector $ \left\langle I \right| $.  For $\lambda \neq I $, the
source term in (\ref{proof}) is also absent, leading to the solution
$\left\langle I \right| J_{12}(t) \left| \lambda \right\rangle = 0 $.
Using this property, we may replace the atomic average
over products of atomic operators with products of atomic averages,
provided that $J_3(t)=J_3(0)$. For example, 
\begin{eqnarray}
\left\langle J_{12}J_{21}\right\rangle  &=&\sum_{\mu}\left\langle vac\right| \otimes \left\langle I\,\right| \,J_{12}\,\left|
\,\mu \right\rangle \,\left\langle \,\mu \right| \,J_{21}\left|
I\right\rangle \otimes \left| vac\right\rangle   \nonumber \\
\  &=&\left\langle \left\langle J_{12}\right\rangle _A\left\langle
J_{21}\right\rangle _A\right\rangle _R.
\end{eqnarray}
Here, $\left\langle {\cal O}\right\rangle _R\equiv
\left\langle vac\right| $ ${\cal O}\left| vac\right\rangle $ denotes
an expectation value over
the reservoir variables. For an
arbitrary moment $g^{pq}$, we have 
\begin{equation}
g^{pq}=\left\langle \left\langle J_{12}\right\rangle _A^p\left\langle
J_{21}\right\rangle _A^q\right\rangle _R.  \label{factorize}
\end{equation}
We note that such a factorization is valid only for an antinormal
ordering of the polarization operators, since $\left\langle I \right|
J_{21}(t) \left| \lambda \right\rangle $ does not vanish in general.

Taking the atomic expectation value of equation (\ref{earlyeq}), we obtain 
\begin{equation}
\frac d{dt}\left\langle J_{12}(t)\right\rangle _A=N\int_0^t\left\langle
J_{12}(t^{^{\prime }})\right\rangle _AG(t-t^{^{\prime }})dt^{^{\prime }}+N%
\eta (t).  \label{blah}
\end{equation}
This is a linear equation that has lost its operator character over the
atomic variables but not over the electromagnetic reservoir, as
evidenced by the presence of
the quantum noise operator, $\eta (t)$. Equation (\ref{blah}) can be
solved by the method of Laplace transforms. The solution has the form, 
\begin{equation}
\left\langle J_{12}(t)\right\rangle _A=D(t)\left\langle
J_{12}(0)\right\rangle _A+N\sum_\lambda C_\lambda (t)a_\lambda (0),
\label{linsoln}
\end{equation}
where, 
\begin{equation}
D(t)={\cal L}^{-1}\left\{ \tilde D(s)\right\} , 
\end{equation}
\begin{equation}
\tilde D(s)= \left[ s - N\tilde G(s)\right] ^{-1} ,  
\label{blah2}
\end{equation}
and
\begin{equation}
C_\lambda (t)={\cal L}^{-1}\left\{ \frac{g_\lambda }{s+i\Delta _\lambda }%
\tilde D(s)\right\} .  \label{convolution}
\end{equation}
Again, ${\cal L}^{-1}$ denotes the inverse Laplace transform.
The Laplace transformation of the memory kernel for an
isotropic band edge, $\tilde G_I(s)$, is
given in equation (\ref{gs}). Despite the fact that $\left\langle
J_{12}(0) \right\rangle_A = 0 $, we retain the first term in equation
(\ref{linsoln}) for later notational convenience

The early-time quantum fluctuations in a superradiant system prevent us from
predicting {\em a priori} the evolution of any single experimental
realization of the atoms. Instead, we can only determine the probability of
a particular trajectory of the collective atomic Bloch vector. In order to
obtain the statistics of a band edge superradiant pulse, we first determine
the statistics of the collective Bloch vector for a set of
identically--prepared systems after each has passed through the early time
regime governed by vacuum fluctuations. The relevant time scale will be
referred to as the quantum to semi-classical evolution {\em crossover time}, 
$t=t_0$. Our approach is to calculate the phase and amplitude distributions
of the polarization at the crossover time quantum mechanically. The
subsequent ($t>t_0$) evolution of the ensemble is then obtained by solving
the semi-classical equations (\ref{semij3}) and (\ref{semij12}) using the
polarization distribution function at $t_0$.  In other words, the
distribution of values 
of $\left\langle J_{12}(t_0) \right\rangle$ obtained from the
early time quantum fluctuations provide the initial conditions for
subsequent, semi--classical evolution. In order to implement this
approach, we must first identify $t_0$ for our system \cite
{{haroche},{haakesr}}. One expects such a transition to occur in the high
atomic inversion regime, $\left\langle J_3(t)\right\rangle \simeq N$.  It is
natural to define
$t_0$ such that
for $t>t_0$ the expectation value of the commutator of the system operators 
$J_{21}(t)$ and $J_{12}(t)$ becomes very small compared to the
expectation value of their product \cite{haroche}. This gives the condition, 
\begin{equation}
\left\langle J_{21}(t)J_{12}(t)\right\rangle \gg \left\langle \left[
J_{21}(t),J_{12}(t)\right] \right\rangle ,\quad t>t_0.  \label{condition}
\end{equation}
Evaluating the above commutator, we have $\left\langle \left[
J_{21}(t),J_{12}(t)\right] \right\rangle =\left\langle J_3(t)\right\rangle $%
, which is equal to $N$ for full atomic inversion. From
(\ref{factorize}) and (%
\ref{linsoln}), we
find that 
\begin{eqnarray}
\left\langle  J_{21}(t)J_{12}(t) \right\rangle &=& \left\langle J_3(t) +
J_{12}(t)J_{21}(t) \right\rangle \nonumber \\
&=& N \left[ 1+N\sum_{\lambda} \left|C_{\lambda} (t) \right|^2\right ]
\nonumber \\
&=& N\left| D(t)\right| ^2,  \label{lincorrelation}
\end{eqnarray}
The last equality is obtained by use of the 
identity $N\sum_\lambda \left| C_\lambda (t)\right| ^2=\left|
D(t)\right|^2-1$, as derived in Appendix B. In free space, $\left| D(t)\right| ^2=e^{N\gamma t}$,
giving the crossover time, $t_0^{free} \simeq 1/N\gamma
$. One can solve
for the crossover time near a band edge, $ t_0^{PBG}$,
computationally. In the isotropic model, for $\delta_c =0$ we find that $%
t_0^{PBG} \simeq 1.24/N^{2/3}\beta _1$. The
crossover time
maintains this $1/N^{2/3}\beta _1$ dependence for $\omega _{21}$
displaced from the
band edge. The corresponding time scale for the anisotropic gap is $%
1/N^2\beta _3$. The build-up of a macroscopic polarization then occurs more
slowly near an isotropic and more quickly near an anisotropic band edge than
in free space.

Using a semi-classical approach, we may write the value of the polarization
at any time $t\geq t_0$ in terms of an amplitude $\kappa $ and a phase $\phi 
$, $\left\langle J_{12}(t)\right\rangle ^{Cl}\equiv J(\kappa ,\phi ,t)$. The
superscript $Cl$ refers to the fact that the expectation value $\left\langle 
{\quad }\right\rangle ^{Cl}$ is taken in the semi-classical regime $t\geq t_0
$. We define $P(\kappa )d\kappa $ as the probability of finding the
amplitude between $\kappa $ and $\kappa +d\kappa $, and $Q(\phi )d\phi $ as
the probability of finding the phase between $\phi $ and $\phi +d\phi $. We
may then write the moments of the macroscopic polarization distribution as 
\begin{eqnarray}
& &\left\langle \left( J_{12}(t)\right) ^p\left( J_{21}(t^{^{\prime }})\right)
^q\right\rangle ^{Cl} \nonumber \\
&\qquad& = \int d\kappa \int d\phi P(\kappa )Q(\phi )\left[
J(\kappa ,\phi ,t)\right] ^p\left[ J^*(\kappa ,\phi ,t^{^{\prime
}})\right] ^q. \nonumber \\
 \label{l}  
\end{eqnarray}
For $t=t_0$, we assume that the polarization has the form $J(\kappa ,\phi
,t)=\kappa e^{i\phi }$, giving for the moments 
\begin{eqnarray}
&&\left\langle \left( J_{12}(t_0)\right) ^p\left( J_{21}(t_0)\right)
^q\right\rangle ^{Cl}\nonumber \\
&\qquad& =\int d\kappa \int d\phi P(\kappa )Q(\phi )\kappa
^{p+q}e^{i(p-q)\phi }.  \label{cl correlation}
\end{eqnarray}
The quantum analogue, $\left\langle {\quad } \right\rangle ^Q $, of (\ref{cl
correlation}) can be written
in the form of equation (\ref{factorize}) evaluated at $t=t_0 $.
Substituting (\ref{linsoln}) and its adjoint into (\ref{factorize})
yields:
\begin{eqnarray}
&&\left\langle \left( J_{12}(t_0)\right) ^p\left( J_{21}(t_0)\right)
^q\right\rangle ^Q \nonumber \\
&=&N^{p+q}\left\langle \left[ \sum_\lambda C_\lambda
(t_0)a_\lambda (0)\right] ^p\left[ \sum_\lambda C_\lambda
^{*}(t_0)a_\lambda ^{\dag }(0)\right] ^q\right\rangle _R. \nonumber \\
& &  \label{blah3}
\end{eqnarray}
As the reservoir expectation value is taken over the the operators
$a_\lambda $
which satisfy a Gaussian probability distribution, Wick's theorem\cite
{louisell} is applied in order to reduce the operator averages of products
of field operators to averages over products of pairs of field operators. We
then have 
\begin{equation}
\left\langle \left( J_{12}(t_0)\right) ^p\left( J_{21}(t_0)\right)
^q\right\rangle ^Q\simeq \delta _{pq}N^pp!\left| D(t_0)\right| ^{2p}.
\label{q correlation}
\end{equation}
This expression has corrections of order $N^{p-1} $, meaning that it
is asymptotically 
valid for large N.  Equating (\ref{cl correlation}) and (\ref{q
correlation}), we solve for the
distributions $P(\kappa )$ and $Q(\phi )$  to obtain the desired initial
polarization distribution for the semi-classical superradiance equations.
The early time distributions for free space and the band edge differ only in
the form of the function $D(t)$, as the above analysis makes no other
distinction between the two cases. Thus in the band edge system, as in free
space, the entire effect of the early time atomic evolution can be
recaptured using the distribution of initial conditions given at $t=t_0$.
The phase of the polarization is given by the relation 
\begin{equation}
\int_0^{2\pi }d\phi e^{i(p-q)\phi }Q\left( \phi \right) =\delta _{p,q}.
\label{poldist}
\end{equation}
This shows that $Q(\phi )$ is uniformly distributed between $0$ and $2\pi $.
The initial polarization amplitude distribution is found from the relation 
\begin{equation}
\int_0^\infty d\kappa P\left( \kappa \right) \kappa ^{2p}=p!\left[ N\left|
D\left( t_0\right) \right| ^2\right] ^p.
\end{equation}
The result is a Gaussian distribution of width $N\left| D\left( t_0\right)
\right| ^2$ centered at zero, 
\begin{equation}
P(\kappa )=\frac 1{\sqrt{\pi N\left| D\left( t_0\right) \right| ^2}}\exp
\left[ \frac{- \kappa ^2}{N\left| D\left( t_0\right) \right| ^2}\right] .
\label{ampldist}
\end{equation}

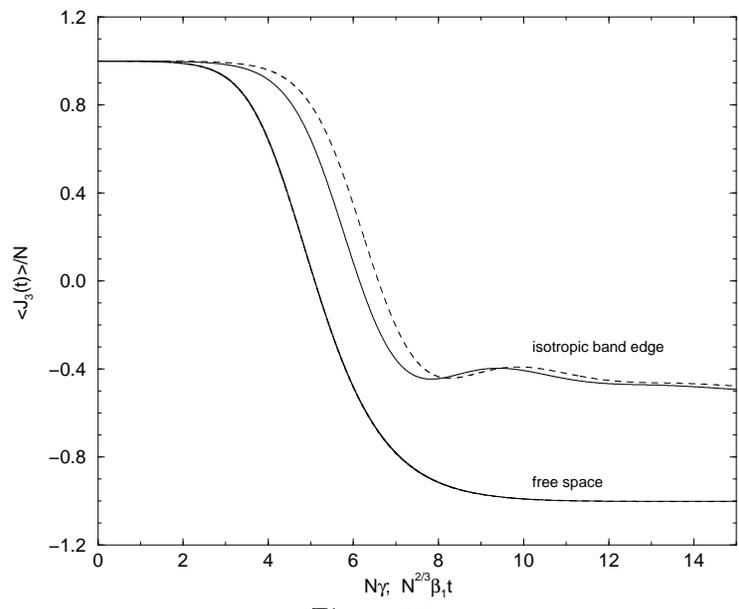
\begin{figure}[tbp]
\caption{Atomic inversion for superradiance driven by vacuum fluctuations in
free space and for an atomic resonant frequency tuned to an isotropic
photonic band edge ($\delta_c=0$). 
Solid lines: result for initial
polarization distribution at $t=0$ for each system; dashed lines:
result for initial
polarization distribution at $t=t_0$ for each system.}
\label{fig9}
\end{figure}

It has been shown via density matrix methods\cite{haroche}
that in free space one may choose the crossover time anywhere in the
inverted regime, the
simplest choice being $t_0=0$. This is due to the absence of temporal
correlations of the reservoir for $t\neq t^{^{\prime }}$.  Figure
\ref{fig9} shows
the ensemble--averaged collective emission in free space and at an isotropic
band edge ($\delta _c=0$) for N=100 atoms. Both the free space and band edge
systems are shown for two choices of initial polarization distribution. The
solid lines correspond to the choice of $t_0=0$ in the amplitude
distribution (\ref{ampldist}) for both free space and the band edge. The
dashed lines correspond to the choice $t_0=t_0^{free}$ and $t_0=t_0^{PBG}$
for the free space and band edge systems respectively. As per equation (\ref
{poldist}), the initial phase of the polarization in all cases is chosen
from a uniform random distribution. In free space, the two choices of
initial conditions yield indistinguishable atomic dynamics. This verifies
that the choice of $t_0$ is unimportant in free space, so long as it is
chosen in the inverted regime. Near a photonic band edge, we see that the
choice of $t_0$ affects the later evolution of the system. In particular, it
affects the onset time for collective emission. It is clear from these
simulations that the details of the non-Markovian evolution in the quantum
regime play a crucial role in the subsequent semi-classical evolution of the
band edge superradiance.  The long--range temporal correlations of the
reservoir require that we treat the vacuum fluctuations explicitly
throughout the quantum evolution of the system. A similar picture
holds in the case of an anisotropic PBG material.  In our anisotropic model,
memory of the initial state is expressed through the Green function (\ref
{greenfn1}). In this case, superradiance is also highly sensitive to
early stage quantum fluctuations.

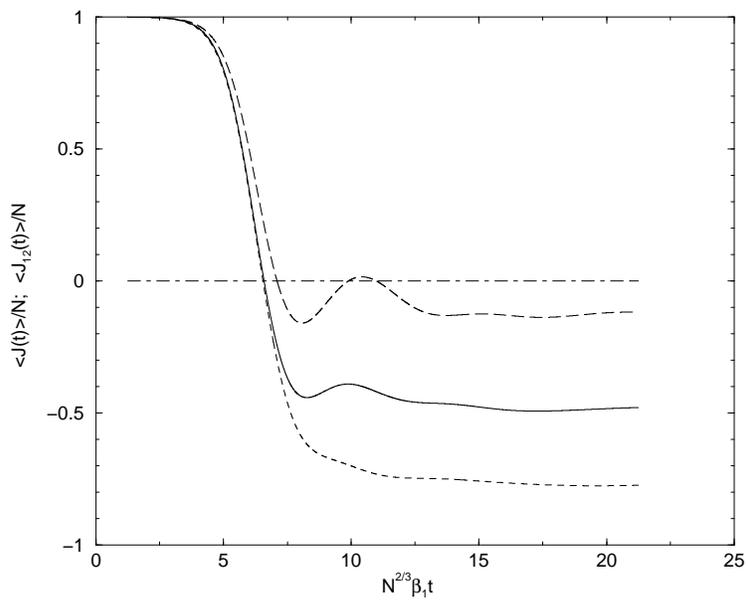
\begin{figure}[tbp]
\caption{Ensemble--averaged atomic inversion, $\left\langle J_3(t)
\right\rangle_{ens}/N $,  and atomic polarization
amplitude, $\left| \left\langle J_{12}(t) \right\rangle_{ens} \right|
/N $  
(dot-dashed line), for a system of $N=100$ atoms near an isotropic
photonic band edge. The ensemble average
is taken over 2000 initial polarization
values. Inversion: long dashed curve, $\delta_c=-.5$; solid line, 
$\delta_c=0$; short dashed line, $\delta_c=.5$.  $\delta_c$
in units of $N^{2/3}\beta_1 $.}
\label{fig10}
\end{figure}

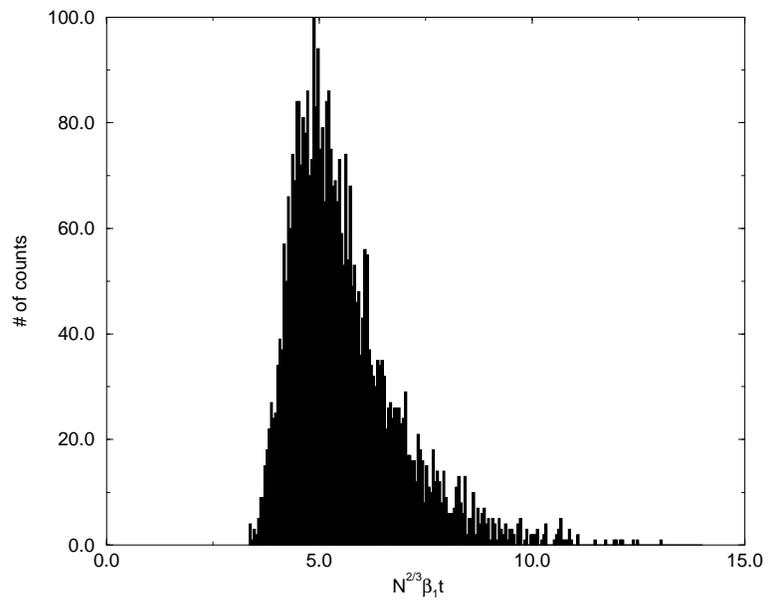
\begin{figure}[tbp]
\caption{Distribution of delay times for a system of 100 atoms at an
isotropic band edge ($\delta_c=0$) for 2000 realizations of the superradiant
system.}
\label{fig11}
\end{figure}

Since ensemble averages of atomic observables are experimentally measurable
quantities, we consider these in some detail. We use the notation $%
\left\langle {\quad }\right\rangle _{ens}$ to denote an ensemble--averaged
quantum expectation value. For illustration, we focus on the $\delta _c=0$
and zero dephasing case for a system of 100 atoms in the isotropic effective
mass model. The extension to non--zero detuning and finite dipole dephasing
follows from the discussion of Section IV.  From Fig. \ref{fig10}, it is
evident that the ensemble exhibits a fractional population inversion in the
steady state. The steady state value of $\left\langle J_3(t)\right\rangle
_{ens}$ for a given atomic detuning is unchanged from the mean field result, 
$\left\langle J_3(t_s)\right\rangle $.  Since the steady state is determined
by the atom--field coupling strength, and not by the dynamics of the system, it is
insensitive to initial conditions. Fluctuations in the excited state atomic
population may be expressed in terms of the delay time for the onset of
superradiant emission, defined as the time at which the system is exactly
half--excited, i.e. $\left\langle J_3\right\rangle =0$. Vacuum fluctuations
result in a distribution of delay times for the ensemble, asymmetrically
centered about a peak value, as pictured in Fig. \ref{fig11}. The delay time
distribution is qualitatively similar to that obtained in free space\cite
{haakesr}. However, the width of the distribution scales with the
relevant time scale for the isotropic and
anisotropic gaps, showing
that, near a photonic
band edge, atomic population fluctuations during light emission can be
reduced from their free 
space value.  Because of the variation in initial conditions, the Rabi
oscillations in $\left\langle J_3(t)\right\rangle $ for the isotropic gap
are much less pronounced than in mean field simulations. The differences in
emission times due to fluctuations cause the ensemble average inversion to
smear out these oscillations.  Therefore, one can no longer directly relate the
amplitude and period of the oscillations to the energies of the collective
dressed states.

More striking is the nature of the ensemble's collective polarization under
the influence of vacuum fluctuations. Figures \ref{fig12} a -- d show the
evolution of the polarization distribution from the initial distribution
given by equations (\ref{poldist}) and (\ref{ampldist}) to the steady state
distribution. Initially, the distribution is sharply peaked about zero. In
the decay region, the polarization amplitude is broadly distributed and has
a random phase. This behavior is reminiscent of the fluctuations of the
order parameter in the vicinity of a phase transition. In the steady
state, the polarization amplitude collapses to a very well--defined
non--zero value. This amplitude is again accompanied by a random phase that
is uniformly distributed between $0$ and $2\pi $. We may interpret our
steady state result in the following manner: A fraction of the photons
emitted near the photonic band edge remain localized in the vicinity of the
atoms, causing both the atomic dipoles and the electromagnetic field to
self--organize into a cooperative steady state. However, vacuum fluctuations
cause this cooperative quantum state to have a random phase, resulting in a
zero ensemble average polarization amplitude, $\left| \left\langle
J_{12}(t)\right\rangle _{ens}\right| =0$, as shown in
Fig. \ref{fig10}.  Measurements of the
degree of first and second order coherence of the electromagnetic field in a
band edge superradiance experiment would provide a probe of the nature of
this self--organized state of photons and atoms near a band edge. Such
measurements would however have to be done by indirect means, as the
localized radiation field in a PBG does not propagate out of the photonic
crystal. We further note that this state - well--defined in amplitude but
with random phase - is similar to the steady state of a conventional laser%
\cite{ssandl} with a well--defined electric field and random phase diffusion.

\section{Simulated Quantum Noise Near a Band Edge}

We have shown that the statistical properties of a band edge superradiant
system can be determined because the collective behavior of the constituent
atoms leads to a semi-classical system evolution, triggered by early-time
quantum fluctuations. However, a seamless quantum description of  band
edge quantum
optical systems is extremely difficult to obtain, due to the
non-Markovian nature of
the atom-field interaction. As a first step,
we introduce a method by which to simulate their evolution
computationally and include the
effects of quantum fluctuations.  Unlike the semi-classical simulations of
section IV which neglected the effect of the quantum noise operator, as $%
\left\langle \eta (t)\right\rangle =\left\langle \eta ^{\dag
}(t)\right\rangle =0$, we propose to replace $\left\langle 
\eta (t)\right\rangle $ in our semi-classical equations by a complex
classical stochastic function with the same mean and two-time
correlation function as its quantum counterpart. This noise function then
simulates the quantum noise in our system throughout the entire system
evolution. We may test the validity of
our simulated noise ansatz for band edge superradiance by comparing the
results obtained to those calculated in Section V.

The classical noise function required to simulate quantum noise near a
photonic band edge involves  a real
stochastic function $\xi (t)$ possessing the underlying temporal
autocorrelation of our non--Markovian quantum noise operator, $\eta
(t) $.  In the effective mass approximation, this means that (see
equations (\ref{greenfn2}) and (\ref{greenfn1})), 
\begin{equation}
\left\langle \xi (t)\xi (t^{^{\prime }})\right\rangle =\frac 1{%
(t-t^{^{\prime }})^{\alpha /2}},  \label{etacorr}
\end{equation}
where again $\alpha = $ 1 and 3 for isotropic and anisotropic band edges
respectively.  Problems in band edge atom--field
dynamics, such as the present superradiant problem, often involve
non-linear equations
under the
influence of colored quantum noise.  It is interesting to note that
nonlinear problems involving
classical colored noise are of considerable interest in classical
statistical physics. In particular, methods for computationally generating
noise satisfying equation (\ref{etacorr}) have been developed in the context
of problems in surface growth and polymer ordering\cite{colornoise}. In what
follows, we use the method first introduced by Rice \cite{rice} and
elaborated on by Billah and Shinozuka \cite{shinozuka} for the
generation of such colored noise. For noise with a power spectrum $P(\omega
)$, defined as the Fourier transform of the autocorrelation function
$\left\langle \xi(t) \xi(t^{\prime}) \right\rangle $, their algorithm gives, 
\begin{eqnarray}
\xi(t) &\simeq& 2 \sum_{n=1}^N\left[ P(\omega _n)\Delta \omega \right]
^{1/2}\cos (\omega _nt+\Phi _n), \nonumber \\
&\qquad& \qquad \quad \qquad \qquad \qquad \qquad n=1,2,...,N,
\label{noisegen} 
\end{eqnarray}
with equality obtained for $N \rightarrow \infty $.  Here, $\omega
_n=n\Delta \omega $, $\Delta \omega =\omega _{\max }/N$, and
$\omega _{\max }$ is a cutoff frequency above which the power spectrum
can be neglected. Each $\Phi _n$ is a random phase uniformly distributed in
the range $\left[ 0,2\pi \right] $ . By use of a particular set of random
phases $\left\{ \Phi _n\right\} $ to generate the noise values at each time
step, we obtain a single ''experimental'' realization of the quantum noise
in our system. Since we cannot predict {\em a priori} the specific form of the
quantum fluctuations in a particular experiment, we again average over many
realizations of the superradiant system, each governed by a different $\xi
(t)$, in order to obtain distributions and ensemble averages of relevant
quantities.  We note that equation (\ref{noisegen}) clearly gives
$\left\langle \xi(t) \right\rangle_{ens} = 0 $, as desired, since the
random $\Phi_n $ cause the ensemble to average to zero.  To show that
(\ref{noisegen}) also gives the correct autocorrelation function, we
write:
\begin{eqnarray}
&\left\langle \right.&   \xi(t) \xi ( t^{\prime} )
\left. \right\rangle  _{ens} \nonumber \\
&=& \left\langle \right.
4\Delta\omega\sum_k \sum_l
\left[ P\left( \omega_k \right) P\left( \omega_l
\right) \right]^{1/2} \cos \left( \omega_k t + \Phi_k \right)
\nonumber \\
&\quad& \times \cos \left
( \omega_l t^{\prime} + \Phi_l \right) \left. \right\rangle_{ens}
 \nonumber \\   
&=& \langle 2 \Delta\omega \sum_k \sum_l \left[ P\left(\omega_k \right)
P \left( \omega_l \right) \right]^{1/2} \nonumber \\
&\quad & \times \left\{ \cos \left( \omega_k t
-\omega_l   t^{\prime}+ \Phi_k - \Phi_l \right)
\right. \nonumber \\
&\quad & + \left. \left. \cos \left( \omega_k t + \omega_l t^{\prime}
+ \Phi_k + \Phi_l\right) \right\} \right\rangle_{ens} \label{autocorr1}\\
&=& \left\langle 2 \Delta \omega \sum_k P \left( \omega_k \right) \cos
\left[ \omega_k \left( t-t^{\prime} \right) \right] \right\rangle_{ens},
\label{autocorr}
\end{eqnarray}
In (\ref{autocorr1}), only the $k=l $ components, in which the random phases
$\Phi_k $ and $\Phi_l $ cancel each other, survive the ensemble
average.  All other terms in
(\ref{autocorr1}) vanish in the ensemble average.  As $N \rightarrow
\infty $, (\ref{autocorr})
becomes the Fourier transform of $P( \omega ) $,
which equals $\left\langle \xi(t) \xi(t^{\prime})
\right\rangle_{ens}$ .  Studies have shown that for values of $N $ as
small as 1000, the desired autocorrelation may be obtained with as
little as $5 \% $ error\cite{shinozuka}, making this a computationally feasible
technique. Furthermore, unlike other methods for the generation of
stochastic functions (see
Refs.\cite{colornoise}), the present method computes the desired function,
$\xi(t) $, 
using only a uniform random distribution of phases $\Phi_k $ as input,
rather than
requiring the computation of a Gaussian
stochastic function as an intermediate step.  This decreases the
likelihood of spurious correlations between our random numbers.
Figure \ref{fig14}  shows the two--time correlation of $\xi (t)$ for $\alpha
=1$, for an ensemble
of 2000 realizations of the noise function generated by the algorithm of
equation (\ref{noisegen}). In this calculation and in the simulations
described below, we chose a power
spectrum $P(\omega ) = \sqrt{\frac{\pi}{2\omega}} $, in order to mimic
the colored vacuum near an isotropic band edge.
 We see good agreement with the correlation
function (\ref{etacorr}). The agreement between our simulations and
the exact correlation function can be significantly improved by
enlarging the size of the ensemble, at the expense of increased
computation time for atom--field simulations.

The ensemble $\left\{ \xi
(t)\right\} $ is used to simulate the effect of vacuum fluctuations in
equations (\ref{j3eq}) and (\ref{j21eq}). Written in terms of the
dimensionless time variable $\tau =N^{2/3}\beta _1t$, these equations
for the isotropic band edge become 
\begin{eqnarray}
&\qquad & \frac{d}{d\tau } \left\langle J_3(\tau )\right\rangle \nonumber \\
&=& -4Re\left\{ \right. e^{-i\pi /4}
\frac{\left\langle J_{21}(\tau )\right\rangle }{\sqrt{\pi }}\int_0^\tau 
\frac{\left\langle J_{21}(\tau ^{^{\prime }})\right\rangle }{\sqrt{\tau
-\tau ^{^{\prime }}}}e^{i\delta _c(\tau -\tau ^{^{\prime }})}d\tau
^{^{\prime }} \nonumber \\
&\quad& + \frac{\left\langle J_{21}(\tau )\right\rangle e^{-i(\pi
/8-\delta _c\tau )}}{\sqrt{N\sqrt{\pi }}}\xi (\tau )\left. \right\}
\end{eqnarray}
\begin{eqnarray}
\frac d{d\tau }\left\langle J_{12}(\tau )\right\rangle &=& e^{-i\pi /4}\frac{%
\left\langle J_3(\tau )\right\rangle }{\sqrt{\pi }}\int_0^\tau \frac{%
\left\langle J_{21}(\tau ^{^{\prime }})\right\rangle }{\sqrt{\tau -\tau
^{^{\prime }}}}e^{i\delta _c(\tau -\tau ^{^{\prime }})}d\tau
^{^{\prime }} \nonumber \\
&\quad& +
\frac{\left\langle J_3(\tau )\right\rangle e^{-i(\pi /8-\delta _c\tau )}}{%
\sqrt{N\sqrt{\pi }}}\xi (\tau ),
\end{eqnarray}
with similar equations for the anisotropic gap. For both models, the noise
term scales as $1/\sqrt{N}$; this is the same dependence of the noise term
on particle number exhibited in free space\cite{polder}. In Fig. \ref{fig15},
we show the average inversion for an ensemble containing 2000
realizations of $\xi (\tau )$ for $N=1000$ and $N=10000$ atoms.  We
find  that our stochastic
simulation scheme gives physical results only for systems of $N > 500$
atoms.  The stochastic simulations show good agreement with the atomic
inversion obtained by the method of Section V.  Other system
properties, such as the ensemble--averaged polarization and the delay
time distribution calculated by the present method also agree well
with the quantum calculations of the previous section.  This suggests
that our stochastic approach may be a valuable tool
in the analysis of band edge atom--field dynamics.

\section{Conclusions}

In this paper, we have treated the collective spontaneous emission of
two--level atoms near a photonic band edge.  An analytic
calculation of the atomic operator dynamics in the case of low atomic
excitation was given.  The results demonstrate novel atomic emission
spectra and show the possibility of reducing atomic population
fluctuations.  This in turn suggests that fluctuations in photon
number are likewise suppressed for light localized near the atoms.
This raises the interesting question of whether squeezed
light\cite{squeezing}, antibunched photons\cite{loudon}, and other
forms of non--classical light may be generated
in a simple manner from band edge atom--field systems.  For an
initially inverted system prepared
with a small macroscopic polarization, a mean field factorization was
applied to the atomic quantum Langevin equations, giving a
semi--classical system evolution.  We found that the atoms exhibit
fractional population trapping  and a macroscopic polarization in the
steady state.  Collective Rabi oscillations of the atomic
population were found, and were attributed to the interference of
strongly--dressed atom--photon states that are repelled from the band
edge, both into and out of the gap.  The degree of photon
localization, the polarization amplitude, and the phase angle of the
polarization in the steady state are all sensitive functions of the
detuning of the atomic resonant frequency from the band edge.  The
steady state atomic properties can thus be controlled by applying a
d.c. Stark shift to the atomic resonant frequency. 

The effect of quantum
fluctuations for high initial excitation of the atoms was included
by distinguishing regimes of quantum and semi--classical collective
atomic evolution.  We found that the early time quantum evolution must
be treated in detail, due to the non--Markovian electromagnetic
reservoir correlations near a band edge.  This is in contrast with
free space, where the atomic system's evolution is insensitive to the
treatment of the full temporal evolution of the early, quantum regime.
Fractional localization of light was shown to persist under the
influence of vacuum fluctuations.  The atomic polarization exhibits a
non--zero amplitude with a randomly distributed phase in the steady
state. This is much
like the steady state of a conventional laser.  Here, such lasing
characteristics are due only to the Bragg scattering of photons back
to the atoms; there is neither external pumping nor a laser cavity in our
system.  The time scales for all dynamical processes, such as
collective emission and the buildup of collective atomic 
polarization are strongly modified from their free space values due
to the singular photon density of states near a photonic band edge.
For an isotropic band edge, the time scales as $N^{2/3}\beta_1 $,
while in the more realistic anisotropic model, time scales as
$N^{2}\beta_3 $.  As a result, collective emission phenomena can occur
more rapidly near a band edge than in free space.  Throughout our
calculations, we have employed an effective mass approximation to the
band edge dispersion.    For materials with a very small PBG, it may
be important to include the effects of both band
edges.  These issues are raised in Appendix A.

We have demonstrated that band edge superradiance possesses many of
the self-organization and coherence properties of a conventional laser.
Furthermore, we have shown the possibility for the generation of
novel emission spectra
and photon statistics.  These results
suggest that a laser operating near a photonic band edge may possess
unusual spectral and stastical properties, as well as a low input
power lasing threshold due to the fractional inversion of the atoms in
the steady state.  It may further be
possible to produce a PBG laser in a bulk material without recourse to
a defect--induced cavity mode.  Lending credence to this
hypothesis, recent 
observations\cite{lawandy} and theoretical
studies\cite{pang} of
lasing from a multiply--scattering random medium with gain have
demonstrated that one may obtain light with the properties of a laser
field in the absence of a cavity.  We are currently investigating the
properties of a
band edge laser, including the effects of pumping and dissipation,
such as the dipolar dephasing modeled in Section IV.  A full
description of the statistics of a band edge laser field will likely require
a non-perturbative master equation (or its equivalent) that exhibits
the non--Markovian nature of the electromagnetic reservoir.
Techniques for treating the atom--field interaction in the absence of
the Born and Markov approximations have recently been obtained
\cite{imamoglu,garraway}.  However, these methods cannot account for
the van Hove singularity in the density of states encountered at a
photonic band edge, nor are they directly applicable to externally
driven atomic systems.  In the absence of a more exact approach, the
band edge  stochastic noise function described in Section VI may be
used to recover many of the results of a full quantum treatment.  

Finally, we note that the steady state atom-field properties described
here are a result of the effect of radiation localized in the vicinity
of the active two--level atoms.  This leads to the question of how to
pump energy into these states, which lie within the forbidden
photonic gap.  One possibility is to couple energy into and/or out of
the system through a third atomic level whose transition energy lies
outside the
gap\cite{sepaper}.  There is also the possibility of transmitting
light into the gap through high intensity ultrashort pulses that
locally distort the nonlinear dielectric constant of the material and thus
allow the propagation of light in the form of solitary waves within
the forbidden
frequency range\cite{neset}.  Such issues must be addressed in order
to fully exploit the very rich possibilities of quantum optical processes
near a photonic band edge. 

\acknowledgements
We are grateful to Dr. Tran Quang for a number of useful discussions.
N.V. acknowledges support from a Walter Sumner Memorial Fellowship.
This work was supported in part by
Photonics Research Ontario, the Natural Sciences and Engineering
Research Council of Canada, and the New Energy and Industrial
Technology Development Organization (NEDO) of Japan.

\appendix 

\section{Calculation of the memory kernel}

We first present the calculation of the memory kernel for the isotropic
model in the effective mass approximation, $G_I(t-t^{\prime })$. Starting
from equation (\ref{gint}) and the isotropic dispersion relation near the
upper band edge, $\omega _k=\omega _c+A(\left| {\bf k}\right| -\left| {\bf k}%
_0\right| )^2$, $G_I(t-t^{\prime })$ can be expressed as 
\begin{equation}
G_I(t-t^{\prime })=\frac{\omega _{21}^2d_{21}^2}{4\hbar \epsilon _0\pi }%
e^{i\delta _c(t-t^{^{\prime }})}\int d\Omega \int_{k_0}^\Lambda dk\frac{%
e^{-iA(k-k_0)^2(t-t^{^{\prime }})}}{\omega _c+A(k-k_0)^2}  \label{isoint}
\end{equation}
Again, $\delta _c=\omega _{21}-\omega _c$ is the detuning of the atomic
resonant frequency from the band edge. $\Lambda =mc/\hbar $ is a cutoff in
the photon wavevector above the electron rest mass. Photons of energy higher
than the electron rest mass probe the relativistic structure of the electron
wave packets of our resonant atoms\cite{bethe}. Because the isotropic model
associates the band edge with a sphere in $k$-space, there is no angular
dependence in the expansion of $\omega _{{\bf k}}$ about the band edge. We
may thus separate out the angular integration over solid angle $\Omega $ in (%
\ref{isoint}). We may also make a stationary phase approximation to the
integral, as the non--exponential part of the integrand will only contribute
significantly to the integral for $k\simeq k_0$. The resulting integral is
\begin{equation}
G_I(t-t^{\prime })\simeq \frac{\omega _{21}^2d_{21}^2}{4\hbar \epsilon _0\pi }%
\frac{k_0}{\omega _c}e^{i\delta _c(t-t^{\prime })}\int_{k_0}^\Lambda
dk\;e^{-iA(k-k_0)^2(t-t^{\prime })}.
\end{equation}
As $\Lambda $ is a large number, we extend the range of integration to
infinity in order to obtain a simple analytic expression for $%
G_I(t-t^{\prime })$\cite{table}:
\begin{eqnarray}
G_I(t-t^{\prime })&=&\frac{\omega _{21}^2\omega _c\sqrt{\omega _{gap}}d_{21}^2%
}{12\hbar \epsilon _0\pi ^{3/2}c^3}\;\frac{e^{-i\left[ \pi /4-\delta
_c(t-t^{\prime })\right] }}{\sqrt{t-t^{\prime }}} \nonumber \\
&=&\beta _1^{3/2}\;\frac{%
e^{-i\left[ \pi /4-\delta _c(t-t^{\prime })\right] }}{\sqrt{t-t^{\prime }}}.
\label{beep}
\end{eqnarray}
Because the relevant frequencies in equation (\ref{beep}) are roughly of the
same order of magnitude near a band edge, we may re--write the prefactor as $%
\beta _1^{3/2}\simeq \omega _{21}^{7/2}d_{21}^2/12\hbar \epsilon _0\pi
^{3/2}c^3$ , in agreement with the value given in Section II.  We
emphasize that the stationary phase method yields the correct
asymptotic behavior for the memory kernel for large $ \left|
t-t^{\prime} \right| $.  At short times, the integral must be
evaluated more precisely using the full photon dispersion relation, as
discussed below.

For an anisotropic band gap model, we must account for the variation in the
magnitude of the band edge wavevector as ${\bf k}$ is rotated throughout the
Brillouin zone. We associate the gap with a specific point in $k$--space
that satisfies the Bragg condition, ${\bf k}={\bf k}_0$. In the effective
mass approximation, the dispersion relation is expanded to second order in k
about this point, $\omega _{{\bf k}}=\omega _c\pm A({\bf k}-{\bf k}_0)^2$.
Making the substitution ${\bf q}=$ ${\bf k}-{\bf k}_0$ and performing the
angular integration, $G_A(t-t^{^{\prime }})$ is expressed as 
\begin{equation}
G_A(t-t^{^{\prime }})=\frac{\omega _{21}^2d_{21}^2}{4\hbar \epsilon _0\pi ^2}%
e^{i\delta _c(t-t^{^{\prime }})}\int_0^\Lambda dq\frac{q^2e^{-iAq^2(t-t^{^{%
\prime }})}}{\omega _c+Aq^2}  \label{anisoint}
\end{equation}
Extending the wavevector integration to infinity, the Green function is\cite
{table} 
\begin{eqnarray}
G_A(t-t^{^{\prime }})&=&\frac{\omega _{21}^2d_{21}^2}{8\hbar \epsilon _0\pi ^2}%
e^{i\delta _c(t-t^{^{\prime }})}\left\{\right. \sqrt{\frac \pi {i\omega
_c(t-t^{^{\prime }})}} \nonumber \\
&-&\frac \pi 2\sqrt{\frac{\omega _c}A}e^{i\omega
_c(t-t^{^{\prime }})}\left[ 1-\Phi \left( \sqrt{i\omega _c(t-t^{^{\prime }})}%
\right) \right] \left. \right\} . \nonumber \\
  \label{ugh}
\end{eqnarray}
$\Phi \left( x\right) $ is the probability integral, $\Phi \left( x\right) =%
\frac 2{\sqrt{\pi }}\int_0^xe^{-t^2}dt$. For $\omega _c(t-t^{^{\prime }})\gg
1$, a condition satisfied for all but the $t^{^{\prime }}\rightarrow t$
limit (as $\omega _c\sim 10^{15}s^{-1}$ for optical transitions),
taking the asymptotic expansion of $\Phi (x)$ to second order gives 
\begin{eqnarray}
G_A(t-t^{^{\prime }})&=&\frac{\omega _{21}^2d_{21}^2}{8\hbar \epsilon _0(\pi
A)^{3/2}\omega _c}\frac{e^{i\left[ \pi /4+\delta _c(t-t^{^{\prime }})\right]
}}{(t-t^{^{\prime }})^{3/2}}, \nonumber \\
&\qquad& \qquad \qquad \qquad \qquad \qquad \quad \omega
_c(t-t^{^{\prime }})\gg 1. \nonumber \\
 \label{ga}
\end{eqnarray}
As $t-t^{^{\prime }}\rightarrow 0_{+}$, (\ref{ugh}) reduces to 
\begin{eqnarray}
G_A(t-t^{^{\prime }})&=&\frac{\omega _{21}^2d_{21}^2}{8\hbar \epsilon _0\pi
^2A^{3/2}}\left[ \sqrt{\frac \pi {i\omega _c(t-t^{^{\prime }})}}-\pi \sqrt{%
\omega _c}\right] , \nonumber \\
&\qquad& \qquad \qquad \qquad \qquad \qquad \qquad \quad t-t^{^{\prime }}\rightarrow
0_{+}. \nonumber \\
 \label{onk}
\end{eqnarray}
$G_A(t-t^{^{\prime }})$ possesses a weak (square root) singularity at $%
t=t^{^{\prime }}$. This is an integrable singularity and can thus be treated
numerically \cite{recipes}.

The effective mass dispersion relation used in the evaluation of $%
G(t-t^{^{\prime }})$ is, strictly speaking, valid only near the photonic
band edge, as it fails to give the required linear photon dispersion
relation for large $\left| k-k_0\right| $ (far away from the gap).
Therefore, the integration of the effective mass dispersion for large
wavevector in (\ref{isoint}) and (\ref{anisoint}) introduces a spurious
contribution to $G(t-t^{^{\prime }})$. This difficulty may be overcome for
an isotropic gap model by using a dispersion relation that has the correct
wavevector dependence for all $k$. The simplest model dispersion with the
correct form is 
\begin{equation}
\omega _k/c=\sqrt{k_0^2+\gamma ^2}+sgn(k-k_0)\sqrt{(k-k_0)^2+\gamma ^2}.
\label{fulldisp}
\end{equation}
The double-valued nature of $\omega _k$ at $k_0$ is made explicit by the
function $sgn(k-k_0)=1$ for $k>k_0$, and $sgn(k-k_0)=-1$ for $k<k_0$. This
gives a gap of width $\omega _{gap}=2\gamma c$, centered about the midgap
frequency $\omega _0=c\sqrt{k_0^2+\gamma ^2}$. Note that equation (\ref
{fulldisp}) gives the correct linear dependence in $k$ for both large
positive and negative $k$, and gives the effective mass dispersion for $%
k\sim k_0$. Like the effective mass model, (\ref{fulldisp}) gives a singular
density of states at the band edges, $\omega _c=\omega _0\pm c\gamma $. The
full dispersion relation allows us to evaluate the influence of both band
edges for arbitrary gap width and atomic resonant frequency. Preliminary
numerical calculations show a stronger reservoir memory effect than
demonstrated in the effective mass model for the isotropic band edge.
This may have a significant effect on theoretical
predictions regarding the atom--field interaction in the vicinity of a PBG.
A further simplification has been made in the anisotropic model, in that we
have not included the dependence of $\omega_{\bf k}$ on the
symmetry of a specific photonic crystal. In a real three dimensional PBG
material, the Bragg condition is satisfied for different values of ${\bf k}$
as the wavevector changes direction in $k$--space. This directional
dependence may lead to a much stronger dependence of the localization of
light on the detuning of $w_{21}$ away from the band edge. The impact on the
atom--field interaction in a PBG of both the full isotropic dispersion model
and more realistic dispersions for three dimensional photonic crystals will be
treated elsewhere.

\section{Evaluation of $\sum_\lambda \left| A_\lambda(\lowercase{t})\right|^2$}

We outline the evaluation of $\sum_\lambda \left|
A_\lambda(t)\right|^2$, used to obtain the low excitation population
fluctuations in Section III, equation (\ref{q}).  A similar procedure
is used to arrive at
equation (\ref{lincorrelation}) in Section V.  Starting from the
Laplace transform
$\tilde A_\lambda(s)$
(equation (\ref{a})), we may use the properties of a convolution of
Laplace transforms in order to write
\begin{equation}
A_{\lambda}(t)=g_{\lambda}\int_0^tdt^{\prime}B(t^{\prime})e^{-i\Delta_\lambda
t}.
\end{equation}   
Therefore, we have
\begin{equation}
\sum_{\lambda} \left| A_{\lambda}(t)\right|^2=\int_0^t dt^{\prime}\int_0^t
dt^{\prime \prime} B(t^{\prime
\prime})B^*(t^{\prime})G(t^{\prime}-t^{\prime \prime}),
\end{equation}
with $G(t-t^{\prime})$ defined as in equation (\ref{gsum}).  We may
re--write this double integral in the form:
\begin{eqnarray}
\sum_{\lambda} \left| A_{\lambda}(t)\right|^2 &=& \int_0^t
dt^{\prime}\int_0^{t^{\prime}}dt^{\prime \prime}B(t^{\prime
\prime})B^*(t^{\prime})G(t^{\prime}-t^{\prime \prime}) \nonumber \\
&\quad& + \int_0^t
dt^{\prime}\int_{t^{\prime}}^t dt^{\prime \prime}B(t^{\prime
\prime})B^*(t^{\prime})G(t^{\prime}-t^{\prime \prime}) \nonumber \\
&=& I_1+I_2,
\end{eqnarray}
where $I_1$ and $I_2$ are the first and second double integrals
respectively.  By changing the order of the integrations in $I_2$, we
obtain
\begin{equation}
I_2=\int_0^t dt^{\prime \prime}\int_0^{t^{\prime \prime}}
dt^{\prime}B(t^{\prime \prime})B^*(t^{\prime})G(t^{\prime}-t^{\prime
\prime})=I_1^*.
\end{equation}
Therefore, $\sum_{\lambda} \left| A_{\lambda}(t)
\right|^2 = 2Re \left \{ I_1 \right \}$, and we need only explicitly
evaluate $I_1$.   The Laplace transform of $B(t)$, $\tilde B(s)$ (equation
(\ref{b})), is equivalent to the Laplace transform of the
equation
\begin{equation}
\frac{d}{dt}B(t)=-N\int_0^t dt^{\prime}G(t-t^{\prime})B(t^{\prime}).
\label{leq}
\end{equation}
Substituting (\ref{leq}) into $I_1$ and its complex conjugate, we  obtain
\begin{eqnarray}
2I_1&=&-\frac{1}{N}\int_0^t dt^{\prime} \frac{d}{dt^{\prime}}  \left|
B(t^{\prime}) 
\right|^2 = \frac{1}{N} \left[ \left| B(0) \right|^2 - \left| B(t) \right|^2
\right]\nonumber \\
&=& \sum_{\lambda} \left| A_{\lambda}(t) \right|^2, \label{aresult}
\end{eqnarray}
as $I_1$ is real.  Substituting the initial condition $\left| B(0)
\right|^2=1$ into (\ref{aresult}) gives the result quoted in
Section III.

\begin{figure}[tbp]
\caption{Atomic polarization distribution for a system of 100 atoms at
an isotropic band edge ($ \delta_c = 0 $), subject to quantum
fluctuations at early
times.  5000 realizations of the superradiant system.  (a) $t={t_0}^{PBG}$;
(b) $t=5$; (c) $t=11$; (d) steady state.  $t$ in
units of $1/N^{2/3}\beta_1$.} 
\label{fig12}
\end{figure}

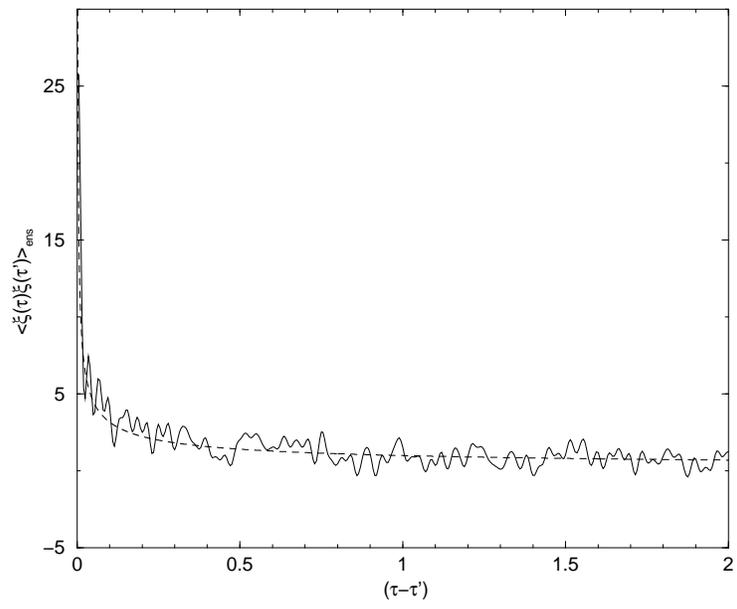
\begin{figure}[tbp]
\caption{Solid line: ensemble averaged autocorrelation function, $\left\langle
\xi (\tau) \xi (\tau^{\prime}) \right\rangle_{ens} $, of the classical colored
noise function $\xi (\tau) $ corresponding to vacuum fluctuations near an
isotropic band edge.  The
dashed line is a plot of the exact autocorrelation function in the
effective mass approximation,
$\left(\tau -\tau^{\prime}\right)^{-1/2} $.} 
\label{fig14}
\end{figure}

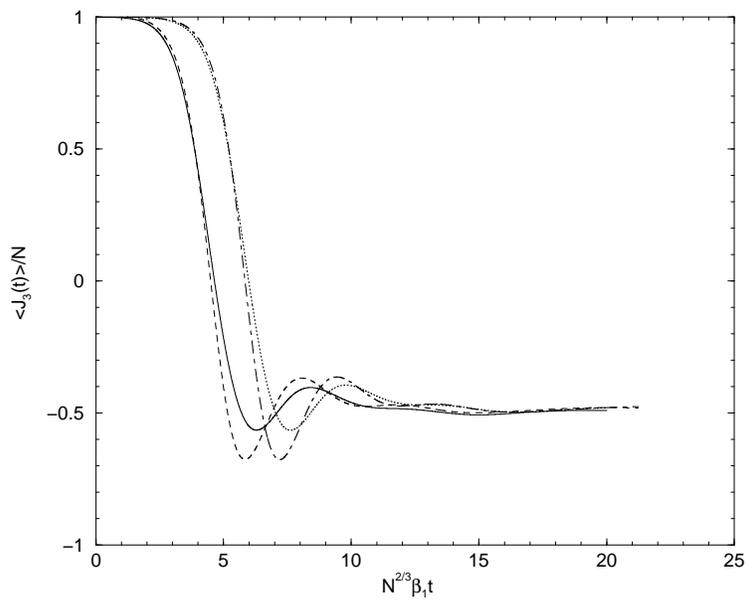
\begin{figure}[tbp]
\caption{Comparison of the ensemble averaged atomic inversion,
$\left\langle J_3(t) \right\rangle_{ens} /N $, at an isotropic band edge
($\delta_c =0$) as
calculated by the methods of Sections V and VI.  2000 realizations of
the superradiant system.  Dashed line and long--short dashed line:
inversion calculated by the method of Section V for $N = $ 1000 and
10000 atoms respectively.  Solid line and dotted line: inversion
calculated using the stochastic function of Section VI for $N = $ 1000
and 10000 atoms respectively.}    
\label{fig15}
\end{figure}

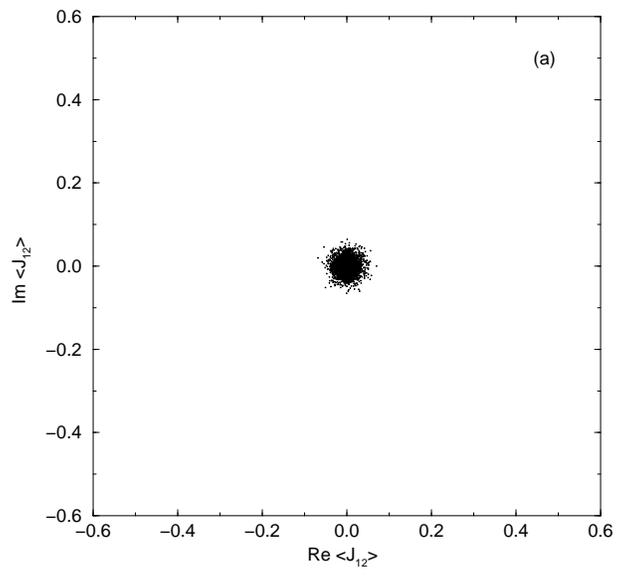
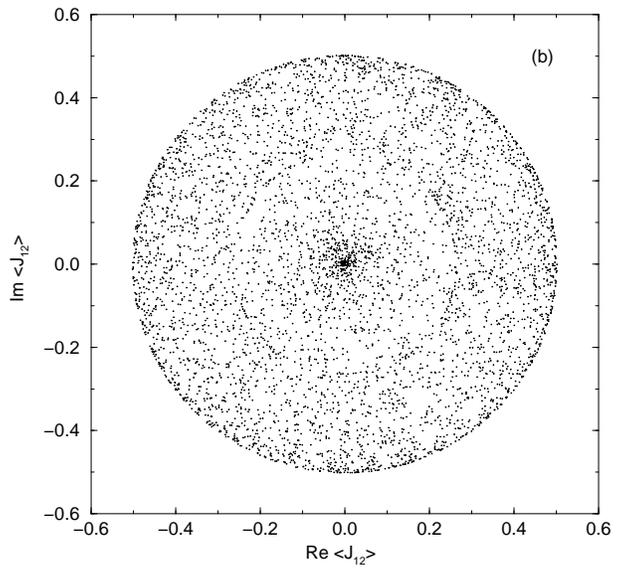
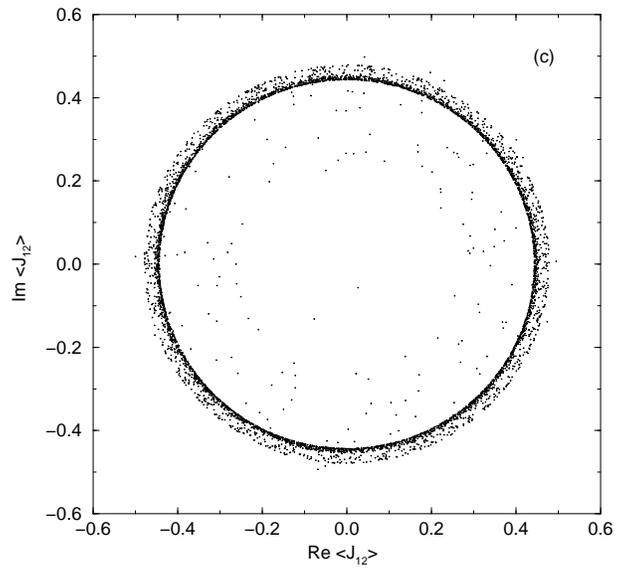
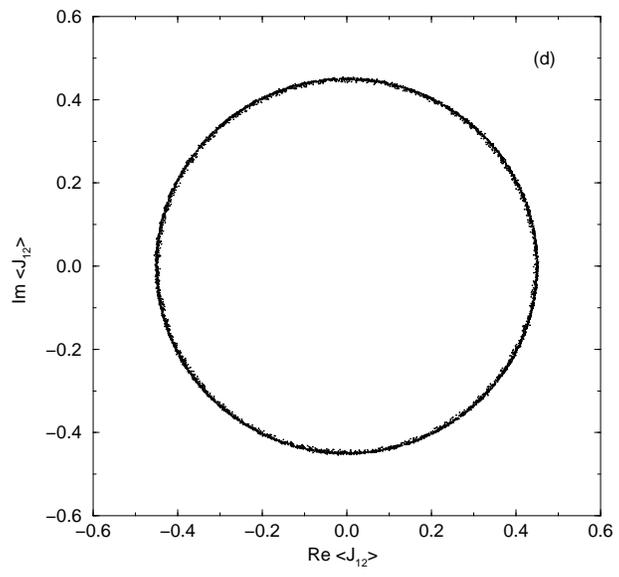
\end{document}